\documentclass[twocolumn]{aastex631}

\received{2026 January 23}
\accepted{2026 April 16}

\usepackage{longtable}
\usepackage{graphics}
\usepackage{epsf}
\usepackage{graphicx}	
\usepackage{amsmath}	
\usepackage{amssymb}	
\usepackage{hyperref}
\usepackage{txfonts}
\usepackage{lipsum}
\usepackage{xcolor}

\shorttitle{Longlived radio outburst of the NLS1 galaxy SDSS\,J1105+1452}
\shortauthors{S. Komossa et al.}

\begin{document}

\title{\large{SDSS\,J110546.07+145202.4: The first long-duration radio changing-look NLS1 galaxy}}

\author[0000-0003-4183-4215]{S. Komossa}
\affiliation{Max-Planck-Institut f{\"u}r Radioastronomie, 
Auf dem H{\"u}gel 69, 53121 Bonn, Germany}

 \author[0000-0002-9961-3661]{D. Grupe}
 \affiliation{Department of Physics, Geology, and Engineering Technology, Northern Kentucky University, Nunn Drive, Highland Heights, KY 41099, USA }

\author[0000-0002-4184-9372]{A. Kraus}
\affiliation{Max-Planck-Institut f{\"u}r Radioastronomie, 
Auf dem H{\"u}gel 69, 53121 Bonn, Germany} 

\author[0000-0002-8186-4753]{P.G. Edwards}
\affiliation{CSIRO Space and Astronomy, Australia Telescope National Facility, PO Box 76, Epping, NSW 1710, Australia} 

\author[0000-0002-0011-6922]{E.F. Kerrison}
\affiliation{Sydney Institute for Astronomy, School of Physics, The University of Sydney, NSW 2006, Australia}
\affiliation{CSIRO Space and Astronomy, Australia Telescope National Facility, PO Box 76, Epping, NSW 1710, Australia}

\author[0000-0002-7329-3209]{K. Rose}
\affiliation{Sydney Institute for Astronomy, School of Physics, The University of Sydney, NSW 2006, Australia}
\affiliation{CSIRO Space and Astronomy, Australia Telescope National Facility, PO Box 76, Epping, NSW 1710, Australia} 

\author[0000-0002-4622-796X]{R. Soria}
\affiliation{INAF-Osservatorio Astrofisico di Torino, Strada Osservatorio 20, I-10025 Pino Torinese, Italy} 
\affiliation{Sydney Institute for Astronomy, School of Physics, The University of Sydney, NSW 2006, Australia}

\author[0000-0003-4341-0029]{T. An}
\affiliation{Department of Astronomy, University of Science and Technology of China, Hefei, Anhui 230026, China} 
\affiliation{State Key Laboratory of Radio Astronomy and Technology, Shanghai Astronomical Observatory, Chinese Academy of Sciences, 80 Nandan Road, Shanghai 200030, People's Republic of China} 

\author[0000-0003-4223-1117]{M.J. Hardcastle}
\affiliation{Department of Physics, Astronomy and Mathematics, University of Hertfordshire, College Lane, Hatfield AL10 9AB, UK} 

\author[0000-0003-1020-1597]{K.\'E. Gab\'anyi}
\affiliation{Department of Astronomy, Institute of Physics and Astronomy, ELTE E\"otv\"os Lor\'and University, P\'azm\'any P\'eter s\'et\'any 1/A, 1117 Budapest, Hungary} 
\affiliation{HUN-REN--ELTE Extragalactic Astrophysics Research Group, ELTE E\"otv\"os Lor\'and University, P\'azm\'any P\'eter s\'et\'any 1/A, 1117 Budapest, Hungary}
\affiliation{Konkoly Observatory, HUN-REN Research Centre for Astronomy and Earth Sciences, Konkoly-Thege Mikl\'os \'ut 15-17, 1121 Budapest, Hungary}
\affiliation{CSFK, MTA Centre of Excellence, Konkoly-Thege Mikl\'os \'ut 15-17, 1121 Budapest, Hungary}

\author[0000-0002-5854-7426]{S. Panda}
\altaffiliation{Gemini Science Fellow}
\affiliation{International Gemini Observatory/NSF NOIRLab, Casilla 603, La Serena, Chile} 

\author[0000-0002-7475-7247]{D.W. Xu}
\affiliation{Key Laboratory of Space Astronomy and Technology, National Astronomical Observatories, Chinese Academy of Sciences, Beijing 100101, People's Republic of China}
\affiliation{School of Astronomy and Space Science, University of Chinese Academy of Sciences, Beijing, People's Republic of China}

\author[0000-0002-6880-4481]{J. Wang}
\affiliation{Key Laboratory of Space Astronomy and Technology, National Astronomical Observatories, Chinese Academy of Sciences, Beijing 100101, People's Republic of China}

\author[0000-0003-3079-1889]{S. Frey}
\affiliation{Konkoly Observatory, HUN-REN Research Centre for Astronomy and Earth Sciences, Konkoly-Thege Mikl\'os \'ut 15-17, 1121 Budapest, Hungary}
\affiliation{CSFK, MTA Centre of Excellence, Konkoly-Thege Mikl\'os \'ut 15-17, 1121 Budapest, Hungary}

\author[0009-0004-6695-5122]{A. Mez\H{o}si}
\affiliation{Department of Astronomy, Institute of Physics and Astronomy, ELTE E\"otv\"os Lor\'and University, P\'azm\'any P\'eter s\'et\'any 1/A, 1117 Budapest, Hungary} 


\begin{abstract}
SDSS\,J110546.07+145202.4 stands out as 
a unique radio changing-look Narrow-line Seyfert 1 (NLS1) galaxy that has brightened dramatically and shows an exceptionally long duration of its ``on" phase. 
We present the first high-frequency radio observations, the first simultaneous radio spectral energy distributions (SEDs),  the first optical--UV--X-ray SEDs, and the first X-ray monitoring and spectroscopy of this recently discovered event.
Importantly for understanding the nature of the outburst, we show that 
the X-ray spectrum is soft with a photon index $\Gamma_{\rm X}=2.5$; line-of-sight absorption and extinction are low or absent; the radio SED is peaked at low frequencies $\sim$2 GHz; and the radio outburst emission is very long-lived ($t>$ 8 yr) and roughly constant.
The softness of the X-ray spectrum, low supermassive black hole (SMBH) mass, and high Eddington ratio all corroborate the optical NLS1 classification. 
We discuss multiple outburst scenarios, including lensing, absorption, a binary SMBH merger, a long-duration giant-star tidal disruption, a newly ignited active galactic nucleus (AGN), 
and an accretion-rate change. While most of them can be either excluded or are deemed too rare and lack positive evidence so far, most or all types of these transients are expected to be detected in ongoing VLA and upcoming SKA surveys. 
SDSS\,J110546.07+145202.4 itself is well explained by an accretion rate change that triggered the powerful radio jet emission.
The low redshift and SMBH mass of this system offer a unique perspective
of the physical processes of radio-jet ignition that are expected to operate in the early Universe around growing SMBHs. 
\end{abstract}


\section{Introduction}

Astrophysical transients shed new light on physics in extreme environments, often in the vicinity of (super)massive black holes.   
In the radio regime, transients probe the early formation and evolution of jets and outflows, and/or the physics of accretion-disk--corona systems, in a regime that is typically different from classical blazars or Galactic binaries, and at short timescales that can be observed conveniently. 
Long-lived extragalactic radio transients on timescales of weeks to years include the afterglows of gamma-ray bursts \citep{Frail1997, Hancock2013}, radio supernovae 
\citep{SukumarAllen1989, Bietenholz2021, Rose2024}, stellar tidal disruption events \citep[TDEs;][]{Bloom2011, Anumarlapudi2024, Li2025}, different forms of unusual AGN outbursts
\citep{Koay2016, Nyland2020, Kunert-Bajraszewska2020, Wolowska2021,
Meyer2025, Birmingham2025}, and 
as yet unidentified transients from the centers of galaxies \citep{Zhang2022, Kunert-Bajraszewska2025, Chen2025}.  
Past and ongoing radio surveys represent excellent databases 
to search for such transient events on timescales of months to decades.

Narrow-line Seyfert 1 (NLS1) galaxies and their higher-luminosity counterparts, NL type 1 quasars (NLQ1s hereafter){\footnote{as common in the literature, we collectively refer to them as NLS1 galaxies, except if noted otherwise}}, are a subgroup of active galactic nuclei (AGN) that are of great interest due to their location at one extreme end of AGN correlation space \citep{Sulentic2000b, Boroson2002, Grupe2004, Xu2012, Marziani2018}. They host low-mass black holes that, on average, accrete close to the Eddington limit. They therefore represent important local cases
\citep[see review by][]{Komossa2008-rev} to study the physics of rapid supermassive black hole (SMBH) growth expected to operate in the early Universe that JWST is now  witnessing \citep{Maiolino2024}. In the radio regime,  NLQ1s and NLS1s have been found to be less frequently radio-loud than their broad-line counterparts (BLS1s and BLQ1s, hereafter). Only $\sim$7\% of NLQ1s are radio-loud \citep{Komossa2006, Chen2022, Rakshit2017}, and only 2.5\% are very radio-loud \citep{Komossa2006}. Nevertheless, like their broad-line counterparts, a small fraction of NLS1s and NLQ1s host relativistic jets and show multiwavelength (MWL) properties similar to blazars \citep{Zhou2007, Yuan2008, Lister2018, Komossa2018}, extending the blazar regime to lower SMBH masses \citep[Fig.\ 4 of][]{Komossa2006} than found in classical blazars and radio-loud quasars. 
Radio variability of NLS1 galaxies as a class is relatively rare; typically much less than a factor of 2 and up to $\sim$5 in rare cases with relativistic jets and at high frequencies $\gtrsim{5}$ GHz \citep{Yuan2008, Angelakis2015, Shao2025}. 

Radio properties of NLS1 galaxies in general, and high-amplitude radio 
outbursts in particular, provide us with important insights into the physics of the jet-disk symbiosis, particle acceleration, 
jet-gas interaction 
in gas-rich and dense circum-nuclear environments
different from classical blazars, 
and feedback processes potentially relevant for galaxy evolution. 

Here, we report follow-up observations and interpretation of the long-lasting, high-amplitude radio outburst of the nearby NLS1 
galaxy SDSS\,J110546.07$+$145202.4 (hereafter SDSS\,J1105$+$1452) 
reported by \citet[][hereafter GKK25]{Gabanyi2025} with an amplitude of a factor $>23$ at 1.4 GHz. 

The MWL properties of SDSS\,J1105+1452 have
been little studied in depth  before. Its optical spectrum, obtained in the course of the Sloan Digital Sky Survey \citep[SDSS;][]{York2000} and included in a number of SDSS large-sample studies \citep[e.g.,][]{Oh2015, Sun2015, Rakshit2017, Paliya2024} is that of a NLS1 galaxy at redshift $z=0.1209$. 
It shows
FWHM(H$\beta_\mathrm{broad}$) $=1498$\,km\,s$^{-1}$ \citep{Sun2015},
strong Fe\,{\sc ii} emission complexes \citep[R4570=0.92,][] {Rakshit2017}, and [O\,{\sc iii}]5007/H$\beta <3$,
and therefore fulfills all NLS1 classification criteria{\footnote{NLS1s are defined by FWHM(H$\beta_\mathrm{broad}$) $< 2000$\,km\,s$^{-1}$, [O\,{\sc iii}]/H$\beta < 3$, and strong Fe\,{\sc ii} emission \citep{OsterbrockPogge1985, Goodrich1989, Veron2001}.  
}}. 
SDSS\,J1105+1452 was detected in X-rays in the ROSAT all-sky survey \citep[RASS;][]{Voges1999}. Based on its absolute B band magnitude of $M_{\rm B} =-20.7$ \citep{Paliya2024} it is in the Seyfert (not the quasar) regime.  
It was identified as radio-emitting NLS1 \citep{Rakshit2017} and detected by LOFAR \citep{Varglund2025}. 
SDSS\,J1105+1452 was radio faint (near the detection limit) in the Faint Images of the Radio Sky at Twenty Centimeters \citep[FIRST,][]{first_white} survey, but has brightened by a factor $\sim$20 since 2017.96 and has turned radio-loud (GKK25). We introduce the term ``radio changing-look'' to refer to high-amplitude radio variability, especially leading to a change between radio-quiet and radio-loud, or vice versa. This is in analogy to the term X-ray changing-look \citep{Matt2003} that refers to high-amplitude absorption changes in X-rays, and optical changing-look \citep{LaMassa2015} that refers to high-amplitude broad line and/or continuum variability.    

 Such radio variability has rarely been observed so far. 
 Among the population of broad-line type 1 quasars, a few have shown remarkable, strong brightening in the radio band and transitioned to radio-loud \citep{Kunert-Bajraszewska2020, Nyland2020}, none as long-lived as SDSS\,J1105+1452, and none in a low-mass AGN or NLS1 galaxy in particular. A few NLS1s exhibit 
 giant-amplitude radio flaring \citep{Jarvela2024}, but these turn on and off on timescales as short as hours to days primarily seen at high frequencies (37 GHz), and deep high-resolution imaging did not show evidence for long-lived jets in the majority of these systems. 

To understand the physics behind the long-lived radio changing-look event of SDSS\,J1105+1452, we have obtained multiple new MWL observations, many of them for the first time.  
This paper is organized as follows. In Sect. 2 we present the Swift X-ray observations. Sect. 3 reports Swift UV-optical observations. In Sect. 4 new radio observations acquired with the Effelsberg 100m telescope and ATCA are presented along with new archival radio observations. Sect. 5 gives a summary of archival IR and optical observations dating back to 2003 and in one case to the 1950s. These are used to constrain the onset of the outburst. In Sect. 6 we apply several methods to determine the SMBH mass of SDSS\,J1105+1452 and its Eddington ratio. Sect. 7 explores the implications of the new MWL observations and provides a discussion of a variety of outburst scenarios. Sect. 8 provides a summary and conclusions and sketches important future observations. 
Throughout this publication, we use a $\Lambda$CDM cosmological model with $H_0=70$\,km\,s$^{-1}$\,Mpc$^{-1}$, $\Omega_\mathrm{m}=0.3$, and $\Omega_\Lambda=0.7$. 
At the redshift of SDSS\,J1105$+$1452, $1''$ corresponds to a projected linear size of $2.18$\,kpc \citep{wright}. 


\section{Swift and ROSAT X-ray observations}

\subsection{Swift XRT}

We have observed SDSS\,J1105+1452 eight times with the Neil Gehrels Swift observatory \citep[Swift hereafter;][]{Gehrels2004} between October and December 2025 to measure its broad-band X-ray spectrum for the first time and search for short-term variability among the Swift data sets and in comparison with the ROSAT detection in the 1990s. An archival Swift observation from October 2023 was added to the analysis. Exposure times range between 0.57 ks and 3.71 ks.

The Swift X-ray telescope \citep[XRT;][]{Burrows2005} was always operating in photon counting mode \citep{Hill2004}.
The XRT data analysis was performed with the latest calibration files and
using the XRTDAS package developed at the ASI Science Data Center (SSDC)
included in the HEASoft package (version No.\ 6.35.1). X-ray count rates were 
determined using the XRT product tool at the Swift
data center in Leicester \citep{Evans2007}. Hardness ratios
$HR = \frac{H-S}{H+S}$
were derived from the event file and source and background counts were selected
in the energy bands S = (0.3--1.0) keV and H = (1.0--10.0) keV. The hardness ratio
was then determined by using the Bayesian Estimation of Hardness Ratios (BEHR) method by
\citet{Park2006}.
Source photons were extracted within a circular region with a radius of 20
detector pixels, where one pixel is equivalent to 2.36$''$.
Background photons were collected in a nearby circular region of radius
236$''$. 
For spectral analysis, source and background spectra were created within XSELECT.
Ancillary Response Files (ARFs) were generated and
the spectra were then binned with 1 count/bin.
We used the most recent response matrix \texttt{{swxpc0to12s6-20210101v015.rmf}}. All uncertainties are reported at the 1$\sigma$ level. 

\begin{figure}[h!]
   \centering
   \includegraphics[width=\hsize]{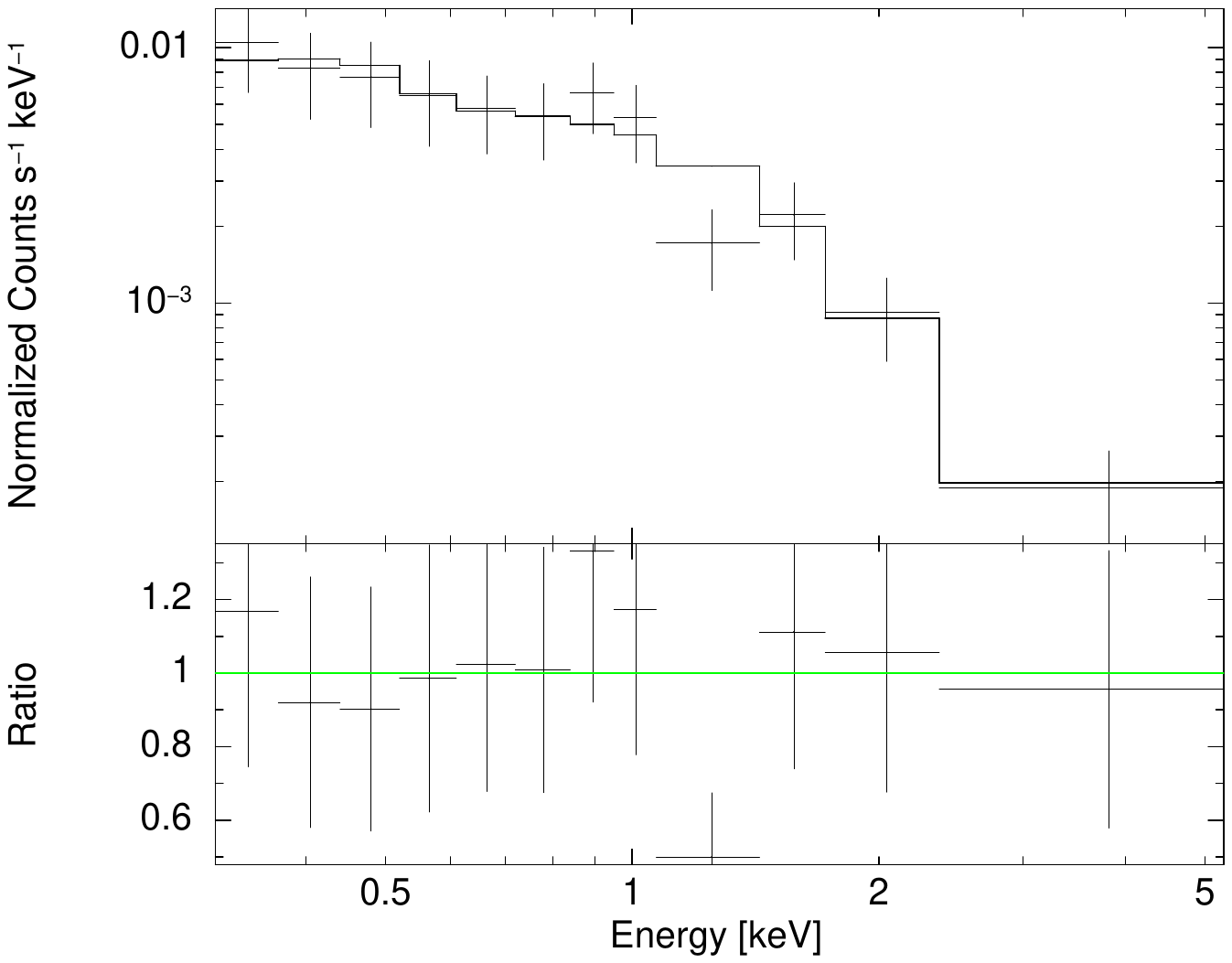}
      \caption{Merged Swift X-ray spectrum of SDSS\,J1105+1452, fit with a single powerlaw model (upper panel) and fit residuals (lower panel). The fit was performed on the unbinned spectrum, the spectrum shown here was binned 
      for visualization purposes.
}
         \label{fig:Xray-spec}
   \end{figure}

The spectral analysis was then performed on the spectra in the 
(0.3--10) keV band using the
software package XSPEC \citep[version 12.15.0;][]{Arnaud1996}.
Since the source was too faint in single
observations, the spectra from all the 2025 observations were merged.
The ARFs were merged and weighted by their individual exposure time with respect
to the total exposure time of the merged spectrum.

\begin{table}[h!]
\caption{X-ray spectral fit results of the combined 2025 XRT data of  SDSS\,J1105+1452. The absorption-corrected flux $f_{\rm int}$ is given in the band (0.3--10) keV.  
}                 
\label{table:X-ray-fit}   
\centering                     
\begin{tabular}{llll}      
\hline\hline               
$N_{\rm H}$ & $\Gamma_{\rm X}$ & $f_{\rm int}$ & $W_{\rm stat}$/d.o.f.  \\
10$^{20}$ cm$^{-2}$ & & $10^{-13}$ erg s$^{-1}$ cm$^{-2}$&  \\ 
\hline
1.32 fixed  & 2.51$\pm{0.16}$  &  $3.0\pm{0.3}$   &  0.86    \\ 
\hline                      
\hline
\end{tabular}
\end{table}

The combined and unbinned spectrum, with a total exposure time of 14.75 ks, was first fit with a single powerlaw model based on W statistics \citep{Cash1979}. Galactic foreground absorption of a column density fixed at $N_{\rm H} = 1.32 \times  10^{20}$ cm$^{-2}$ \citep{HI4PI2016} was included and modeled with TBABS \citep{Wilms2000}.  This gives $\Gamma_{\rm X}$ = 2.51$\pm{0.16}$ (Fig.~\ref{fig:Xray-spec}, 
Tab.~\ref{table:X-ray-fit}). 
Next, the fit was repeated with $N_{\rm H}$ treated as an additional free parameter. 
 In that case the central value for the absorption slightly underpredicts the Galactic value, but within the uncertainties it is unconstrained between 0 and $3.1 \times 10^{20}$ cm$^{-2}$.  
Therefore, there is no evidence for excess absorption intrinsic to the galaxy.  
The fit with fixed Galactic absorption results in an absorbed (0.3--10) keV X-ray flux $f_{\rm abs} = 
(2.8\pm{0.3}) \times 10^{-13}$ erg\,cm$^{-2}$\,s$^{-1}$, an unabsorbed flux of $f=3.0 \times 10^{-13}$ erg\,cm$^{-2}$\,s$^{-1}$, and a luminosity of $1.1 \times 10^{43}$ erg\,s$^{-1}$.  

\subsection{Previous X-ray observations}

SDSS\,J1105+1452 has previously been detected in the ROSAT all-sky survey \citep{Truemper1982} in 1990 with a ROSAT PSPC count rate of 0.087 cts/s in the energy band (0.1--2.4) keV \citep{Voges1999, Anderson2007}. 
Assuming the same spectral model as measured with Swift and Galactic absorption, this corresponds to an absorbed 0.3--10 keV flux of $f_{\rm abs, 1990} = 6.0 \times 10^{-13}$ erg\,cm$^{-2}$\,s$^{-1}$. 
This implies long-term variability by a factor of 2 within 3.5 decades. 

\begin{table*}[]
\caption{Swift UVOT observations of  SDSS\,J1105+1452. The UVOT fluxes, in 
10$^{-13}$ erg\,s$^{-1}$\,cm$^{-2}$, are corrected for Galactic extinction.  
}                
\label{table:Swift-data}    
\centering                        
\begin{tabular}{llllllll}     
\hline\hline               
epoch & UVOT exposure time & $f_{\rm V}$ & $f_{\rm B}$ & $f_{\rm U}$ & $f_{\rm W1}$ & $f_{\rm M2}$ & $f_{\rm W2}$ \\   
      &  s                &  &  &  &   &  &  \\ 
\hline                      
2023-10-30 & 2404 &  $19.3\pm3.9$ & $12.6\pm2.3$ & $11.2\pm1.3$ & $7.8\pm0.7$ & $10.2\pm0.7$ & $10.2\pm0.7$ \\  
2025-10-22 & 1601 & $24.1\pm4.3$ & $17.4\pm2.6$ & $8.9\pm1.5$ & $12.0\pm1.2$ & $13.7\pm1.3$ & $12.9\pm0.9$ \\ 
2025-10-27 & 552 & -- & $19.8\pm2.9$ & $10.6\pm0.5$ & $10.7\pm1.0$ &  -- &   --\\ 
2025-10-30 & 1341 & $22.2\pm4.5$ & $14.2\pm2.4$ & $11.7\pm1.5$ &  $8.2\pm1.1$ & $12.1\pm1.2$ & $11.2\pm0.9$ \\
2025-11-01 & 1649 & $25.2\pm4.0$ & $15.6\pm2.1$ & $12.0\pm1.3$ & $10.5\pm1.0$ & $11.2\pm1.1$ & $12.4\pm0.8$ \\
2025-11-06 & 1733 & $20.6\pm3.3$ & $12.7\pm2.0$ & $11.3\pm1.3$ & $9.5\pm0.9$ & $12.5\pm1.2$ & $11.4\pm0.9$ \\
2025-11-29 & 2319 & $22.0\pm3.0$ & $13.8\pm1.5$ & $12.01\pm1.0$ & $10.5\pm0.9$ & $11.1\pm1.3$ & $13.0\pm0.7$ \\
2025-12-06 & 1774 & $25.9\pm3.0$ & $18.6\pm1.8$ & $11.8\pm1.0$ & $9.6\pm0.8$ & $10.2\pm1.1$ & $9.0\pm0.8$ \\
2025-12-13 & 3710  & $22.2\pm1.9$ & $15.0\pm1.1$ & $11.6\pm0.8$ & $10.9\pm0.7$ & $11.1\pm0.8$ & $11.7\pm0.7$ \\ 
\hline
\hline
\end{tabular}
\end{table*}

\section{Swift UVOT observations}

We also observed SDSS\,J1105+1452 with the Swift UV–optical telescope \citep[UVOT;][]{Roming2005} in all six optical and UV photometric bands with filters V (5468 Å), B (4392 Å), U (3465 Å), UVW1 (2600 Å), UVM2 (2246 Å), and UVW2 (1928 Å). Values in brackets represent the filters' central wavelengths \citep{Poole2008}.

UVOT observations were performed on the same dates as the XRT to measure the UV-optical spectral energy distribution (SED) and search for variability.
Exposure times for the UVOT are in the same range as the XRT observations 
(Tab.~\ref{table:Swift-data}). The UVOT filters V:B:U:W1:M2:W2 are nominally observed with a ratio of 1:1:1:2:3:4 of the total exposure time, respectively.

\begin{figure}[h!]
   \centering
   \includegraphics[clip, trim=1.5cm 5.3cm 1.0cm 4.1cm, width=8.5cm]{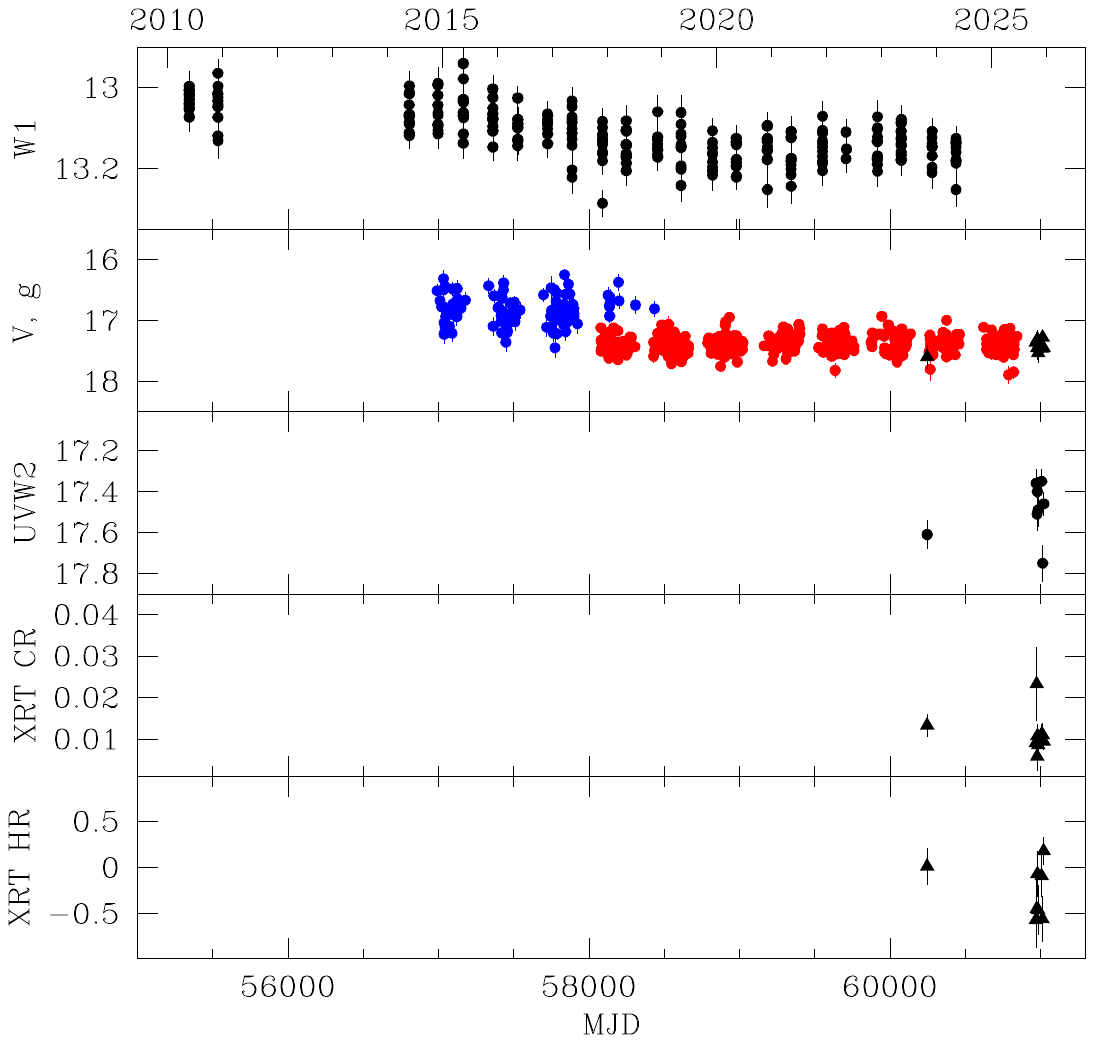}
      \caption{Long-term MWL light curve of SDSS\,J1105+1452 since 2010. Upper panel: WISE and NeoWISE W1 magnitudes, 2nd panel: ASAS-SN V (blue) and g (red) magnitudes (only detections are shown, no upper limits) and Swift V magnitude (black), 3rd panel: Swift UVW2 magnitude, 4th panel: XRT count rate, and 5th panel: X-ray hardness ratio. The two POSS data points between 1949--1965 and 1985--2000, and the SDSS photometry of 2003 (Sect. 5.1) are not displayed.
      }
         \label{fig:MWL-longlight}
   \end{figure}

In each UVOT filter, the observations were first coadded using the tool
\texttt{uvotimsum}. Source counts in all filters were then extracted in a circular
region with a radius of 5$''$ centered on SDSS\,J1105+1452. The background was selected in
a nearby region of 20$''$ radius. Using \texttt{uvotsource},
the background-corrected counts were
then converted into VEGA magnitudes and into fluxes based on the latest
calibration  \citep{Poole2008,Breeveld2010}.
The recently released \texttt{caldb} update, version 2024-02-01, was used. 
All fluxes are reported as flux density multiplied by the central frequency
of the corresponding filter.
A correction of the UVOT data for Galactic reddening was carried out assuming
$E_{B-V} = 0.011$ mag \citep{Schlegel1990} and using a correction factor in
each filter according to Equation (2) of \citet{Roming2005} and adopting the reddening
curves of \citet{Cardelli1989}.
Within the uncertainties, the optical--UV emission of SDSS\,J1105+1452 is essentially constant (Fig.~\ref{fig:MWL-longlight}).  

\section{Radio observations}

\subsection{New Effelsberg observations}
In order to constrain the duration of the outburst and to measure the radio SED for the first time beyond a few GHz, new radio observations of SDSS\,J1105+1452 were obtained at the Effelsberg 100\,m telescope based on a DDT proposal.
Data acquired on 2025 August 4 were already reported in GKK25. Further observations were carried out between August 2025 and December 2025 (program-id 92-25), at frequencies between 4.85 and 16.75 GHz. A simultaneous SED of SDSS\,J1105+1452 between 4.85 and 16.75 GHz was measured for the first time in 2025 September.
Higher-frequency observations were attempted, but were lost due to bad weather. 

Observations were scheduled along with the target OJ~287 from one of our other observing programs \citep[program-id 81-24;][]{Komossa2023}.
These use the same receivers, and the two targets have similar sky locations. This approach allowed us to save observing time by sharing the calibrator sources and set-up.

Standard methods of data acquisition and data reduction \citep{Kraus2003} were followed.
The cross-scan method was used to obtain the radio data. In the cross-scans, the telescope was moved in two perpendicular directions, azimuth and
elevation, with the position of SDSS\,J1105+1452 at the center of the scans. 
Due to the weakness of the source, at least 16 sub-scans (eight per scanning
direction) were performed. All sub-scans of one direction were averaged to 
increase the SNR, before, in a first analysis step, a Gaussian profile was fit to the
data. In some instances, bad sub-scans (e.g. due to RFI) were identified and excluded
from further analysis before the averaging process. 
After a correction for small pointing errors of the telescope, corrections for the opacity of the atmosphere were applied as well as for the
gain-elevation effect (the change of sensitivity with elevation). Finally, absolute flux calibration was achieved by comparing the observed antenna
temperatures with the expected flux densities of calibrator sources like 3C\,286.
The measurement uncertainties are based on the errors resulting from the least
squares fit of the Gaussian profiles and statistical errors from averaging of
the data. These errors are propagated throughout the data reduction process
and combined with a final contribution which reflects the apparent residual fluctuations of the calibrators. 

The measurements (Tab.~\ref{table:radioflux}) show that the high-state is still ongoing and has lasted at least 8 years. The radio flux density at 4.85 GHz remains at $\geq$30 mJy.
The radio spectrum of SDSS\,J1105+1452 declines steeply toward high frequencies. 
The radio emission is roughly constant. 
The monochromatic luminosity at 4.85 GHz is $L_{\rm 4.85\,GHz} = 7 \times 10^{40}$ erg\,s$^{-1}$, and the ratio \citep{Terashima2003} $\log R_{\rm X} = \log L_{\rm 4.85\,GHz}/L_{\rm{(2-10)keV}} = -1.6$ (Tab. \ref{table:radio-summary}).      

\subsection{ATCA observations} 
SDSS\,J1105+1452 was observed with the Australia Telescope
Compact Array (ATCA) on 2025 December 21 between 21:30 and 22:30 UT
 and on January 7 between 14:00 and 15:00 UT (observation code CX601).
The ATCA, which consists of six 22\,m diameter dishes,
was in the east-west 6D array configuration at the time, with a shortest baseline of 77\,m
and a longest baseline of 5878\,m.
Observations were conducted in the 4\,cm and 15\,mm bands, where
the new BIGCAT backend provides 4\,$\times$\,1.92 GHz of bandwidth.
The center frequencies of the four bands were
5.248, 7.168, 9.088, and 11.008 GHz in the 4\,cm band, and
16.960, 18.880, 20.800, 22.720 GHz in the 15\,mm band.
Ten minute observations of PKS B1934$-$638 were made in both
bands for primary flux density calibration.
Ten minute observations of SDSS\,J1105+1452 were bracketed by
two-minute scans on PKS 1055+018 for phase calibration.

BIGCAT data are recorded in ASDM (ALMA Science Data Model) format.
At the time of these observations, an Observatory-developed bespoke version of CASA was
used to convert the data to FITS format. Data reduction was then carried
out in \texttt{Miriad} \citep{Sault1995}. Flux densities were measured in both the $uv-$ and image plane, using a combination of the \texttt{uvfmeas} routine, and imaging with \texttt{robust=0.5}.

SDSS\,J1105+1452 is still in its high-state during the latest measurements of January 2026 (Tab. \ref{table:radioflux}). It is not detected at frequencies $>$18 GHz that are measured with ATCA for the first time, with upper limits of $<$4 mJy. 
The spectral index $\alpha$, defined as in $f_{\rm \nu} \propto \nu^{\alpha}$, is $\alpha_{\rm thin} = -1.3\pm{0.3}$ in the 4\,cm band for the observations between December and January.  

\subsection{Archival radio observations}

 \subsubsection{LOFAR observations}

SDSS\,J1105+1452 was observed seven times at 144 MHz between 2018 and 2024 as part of observations for the third data release (DR3) of the  Low Frequency Array (LOFAR) wide-area survey of the northern sky, LoTSS \citep{Shimwell2026}. The source is clearly detected at a flux density of $\sim 2$ mJy in the LoTSS DR3 mosaics. Because of the survey tiling strategy of LoTSS, many of the individual 8-h observations that cover the source see it only at a large distance from the pointing center and therefore do not allow a good-quality flux density measurement. However, the source is only 0.45 degrees from the pointing center of field P166+15 which was observed for 8 h on 2018 June 12, and this allows a measurement of $2 \pm 0.2$ mJy at this epoch. The other available observations do not provide any statistically significant detections, but are consistent with this flux density and allow us to rule out a strong increase in flux by the end of the observing period (e.g. the flux density is $<3$ mJy at 5-sigma confidence in the observation of field P163+17 on 2024 February 26).

\subsubsection{ASKAP observations}  
We searched the CSIRO ASKAP Science Data Archive \citep[CASDA;][]{Huynh2020} for significant  ($\ge 5\times RMS$) detections in ASKAP catalogues within 5$''$ of SDSS\,J1105+1452's position. We identified several detections from the Rapid ASKAP Continuum Survey \citep[RACS;][]{Hale2021, Duchesne2024,Duchesne2025} at frequencies between 0.944 and 1.368 GHz and the First Large Absorption Survey in H\,{\sc i} \citep[FLASH;][]{Yoon2025} at 0.856 GHz. Flux density measurements were extracted using the \texttt{Selavy} 
source-finding method implemented in the ASKAP processing pipeline; see \cite{Whiting2012} for details. For conservative flux uncertainties, we took  a quadrature sum of the image $RMS$,  the fitted \texttt{Selavy} peak flux error, and a 6\% flux scaling error. Results show that the source was in the high-state during all these observations taken between 2023 and 2025
(Tab.~\ref{table:radioflux}). 

\begin{table*}
\caption{Previous and new radio observations of SDSS\,J1105+1452.}        
\label{table:radioflux}    
\centering                       
\begin{tabular}{lllllcl}      
\hline\hline               
epoch & MJD & survey/ & frequency & integrated flux density & beam size & references \\   
     &  & telescope       & GHz &  mJy & arcsec & \\
\hline            
1986-05 &  & GB6 & $4.85$ & $\lesssim 22$ & $216 \times 204$ & \cite{GB6}\\
1994-06 &     & NVSS & $1.4$ &  $\lesssim 2.5$& $45$ &\cite{nvss} \\ 
1999-12 & & FIRST & $1.4$ &  $1.4 \pm 0.1$ & $5.4$ & \cite{Helfand_first}\\   
2014-03-08 & 56724 & GLEAM & 0.2 & $<45$ & & \cite{Hurley-Walker2017} \\ 
2017-12-18 & 58105  & VLASS & $3$ & $39.0 \pm 0.4$ & $2.5$ &  GKK25 \\ 
2020-04-30 & 58969  & RACS-low &  $0.889$  &   $32.1 \pm 2.8$ & 25 & \cite{Hale2021}\\ 
2020-08-18 & 59079 & VLASS - SE & $3$ & $40.4 \pm 0.4$ & $2.5$ &  GKK25\\
2021-01-04 & 59218 & RACS-mid & $1.367$ &  $32.4 \pm 2.0$ & $11.2 \times 9.3$ & \cite{Duchesne2024}\\
2021-12-31 & 59579 & RACS-high & $1.655$ & $42.7 \pm 4.3$ & $11.9 \times 8.1$ & \cite{Duchesne2025} \\
2023-01-16 & 59960 & VLASS & $3$ &  $43.1 \pm 0.5$ & $2.5$ & GKK25\\ 
2025-08-04 & 60891 & Effelsberg & $4.95$ &  $32 \pm 2 $ &$144$ &  GKK25 \\
 & 60891  & Effelsberg & $6.75$ & $35 \pm 2 $ & $84$ &  GKK25 \\
\hline                                  
2018-06-12 & 58281   & LoTSS & 0.144   & $2.0 \pm 0.2$     &  6  & this paper \\
2023-12-29 & 60307 & RACS-low & $0.944$ & $36.1 \pm 2.1$ &  $14.6 \times 12.1$ & this paper \\ 
2024-01-25 & 60334  & RACS-low & $0.944$ & $44.5 \pm 2.5$ &  $14.7 \times 12.8$& this paper \\
2024-10-31 & 60614 & RACS-mid & $1.368$ & $39.9 \pm 2.3$ & $12.8 \times 8.8$  & this paper \\                               
2025-08-31 & 60918 & Effelsberg & $4.85$ & $30 \pm 5 $ & 145  & this paper \\ 
2025-09-14 & 60932  & Effelsberg & $4.85$ & $38 \pm 2$ & 145 & this paper \\
       & 60932 & Effelsberg & $10.45$ & $17 \pm 2$ & 67 & this paper \\
     & 60932 & Effelsberg & $14.25$ & $19 \pm 3$ & 53 & this paper \\
      & 60932 & Effelsberg & $16.75$ & $18 \pm 2$ & 44 & this paper \\
2025-10-25 & 60973
& FLASH & $0.856$ & $28.8 \pm 1.7^*$ &  $23.1 \times 13.2$ & $^*$uncertain; see Sect. 4.3.3 \\ 
2025-11-04 & 60983  & Effelsberg & $14.25$ & $15 \pm 3$  & 53& this paper \\
      & 60983 & Effelsberg & $16.75$ &  $12 \pm 3$ & 44 & this paper \\
2025-12-21 & 61030 & ATCA & 5.248 &  $26 \pm 4$ & $458 \times 1.4$ & this paper\\
           & 61030 & ATCA & 7.168 &  $23 \pm 4$ & $339 \times 1.0$ & this paper \\
           & 61030 & ATCA & 9.088 &  $13 \pm 4$ & $268 \times 0.8$ & this paper \\
           & 61030 & ATCA & 11.008 &  $8 \pm 3$ & $224 \times 0.6$ & this paper \\
           & 61030 & ATCA & 16.960 &  $4 \pm 2$ & $112 \times 0.5$ & this paper \\
           & 61030 & ATCA & 18.880 &  $<4$ & $102 \times 0.4$ & this paper \\
           & 61030 & ATCA & 20.800 &  $<4$ & $94 \times 0.4$ & this paper \\ 
           & 61030 & ATCA & 22.720 &  $<4$ & $85 \times 0.4$ & this paper \\
2025-12-22 & 61031 & Effelsberg & 4.83 & $32 \pm 2$ & 145 & this paper \\
 & 61031 & Effelsberg & 6.67 & $30 \pm 3$ & 84 & this paper \\ 
 & 61031 & Effelsberg & 10.45 & $18 \pm 2$ & 67 & this paper \\ 
2026-01-07 & 61047 & ATCA & 5.248 &  $31 \pm 4 $ & $195 \times 2.1$ & this paper\\
           & 61047 & ATCA & 7.168 &  $24 \pm 3 $ & $143 \times 1.6$ & this paper \\
           & 61047 & ATCA & 9.088 &  $19 \pm 2$ & $ 114\times 1.2$ & this paper \\
           & 61047 & ATCA & 11.008 & $15 \pm 2$ & $ 93\times 1.0 $ & this paper \\
           & 61047 & ATCA & 16.960 & $2 \pm 1$ & $63 \times 0.6$ & this paper \\
           & 61047 & ATCA & 18.880 &  $<3$ & $57 \times 0.5$ & this paper \\
           & 61047 & ATCA & 20.800 &  $<4$ & $52 \times 0.5$ & this paper \\ 
           & 61047 & ATCA & 22.720 &  $<4$ & $47 \times 0.4$ & this paper \\
 
\hline
\hline 
\end{tabular}
\end{table*}

\begin{figure}[h!]
   \centering
   \includegraphics[clip, trim=0.7cm 1.5cm 3.5cm 0.2cm, angle=-90, width=8.5cm]{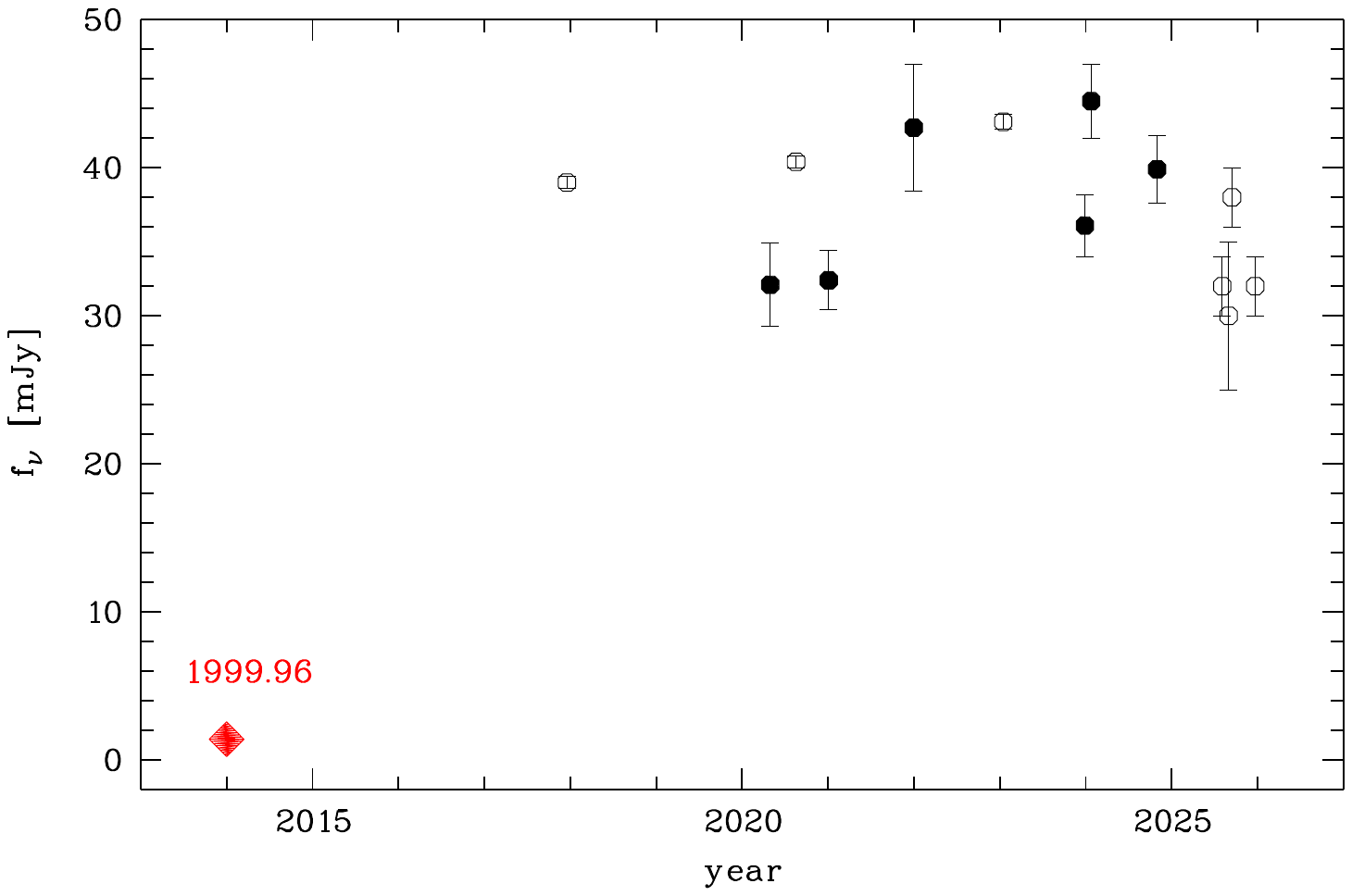}
      \caption{Radio light curve of SDSS\,J1105$+$1452. The leftmost red diamond represents the FIRST low-state data point of 1999.96 (not to scale on the time axis but added for visualization purposes). The more recent high-state data points at all frequencies between 0.8 and 1.7 GHz are shown as filled black circles, and the high-state data points at frequencies between 3 and 5 GHz are represented with open circles. The high-state radio emission since December 2017 has been roughly constant at fixed frequency and already lasted for $>$8 yr. 
}
         \label{fig:lc}
   \end{figure}

\subsubsection{Further archival data}

Tab.~\ref{table:radioflux} also lists additional radio observations from GKK25. These are based on the catalogs of the Green Bank $6$\,cm survey \citep{GB6}, the NRAO VLA Sky Survey \citep[NVSS, ][]{nvss}, the FIRST survey \citep[][]{first_white}, 
the Very Large Array Sky Survey \citep[VLASS, ][]{vlass_lacy} and the RACS \citep{McConnell2020}. 
In addition, we have obtained an upper limit on the 200 MHz flux density 
from the GaLactic and Extragalactic All-sky Murchison Widefield Array (GLEAM) survey image archive \citep{Hurley-Walker2017} of $<45$ mJy. 
SDSS\,J1105+1452 was not detected in the 1980s and 1990s in GB6 and NVSS, and was faint in FIRST (1.4 mJy){\footnote{The GB6, NVSS, and FIRST survey observations extended over a period of months to years and the mid-epochs were assigned as observation dates in the literature (Tab. \ref{table:radioflux}); 1999.96 in the case of FIRST \citep{Helfand_first}.}}. It has been in the radio high-state since the VLASS observation in 2017.96 
(Fig.~\ref{fig:lc}).  

Overall, the high-state radio emission is roughly constant at a $\sim$10\% uncertainty level, with the exception of the FLASH measurement (field SB78421), flagged uncertain in the archive database. According to the validation report{\footnote{\url{ https://ingest.pawsey.org.au/casda-support-files/validation/ASKAP/78421/AS209/validation_image.i.FLASH_667.SB78421.cont.taylor.0.restored.conv/index.html}}}, flux densities in the whole field are systematically low by $\sim$13\%.
The powerlaw spectral index $\alpha_{\rm thick}$ was determined for the low-frequency 
part of the radio SED  
based on the three RACS-low measurements and LoTSS between 0.9--0.14 GHz. This gives $\alpha_{\rm thick} = 1.6 \pm 0.1$.

\begin{table}  
\caption{Summary of the radio properties of SDSS\,J1105+1452. Column (1): Observed rise timescale $t_{\rm rise, radio}$ in the radio band based on the FIRST radio low-state and the first VLASS high-state detection. 
Column (2) Duration $t_{\rm high-state}$ of the high-state. Column (3): Amplitude $A_{\rm{1.4\,GHz}}$ of variability at 1.4 GHz. Column (4): monochromatic 4.85 GHz high-state luminosity $L_{\rm{4.85\,GHz}}$. Column (5): Log of the ratio $R_{\rm X}$ of $L_{\rm{4.85\,GHz}}$ over $L_{\rm{(2-10)\,keV}}$.   
}                 
\label{table:radio-summary}   
\begin{tabular}{ccccc}      
\hline
\hline               
$t_{\rm rise, radio}$ & $t_{\rm high-state}$ & $A_{\rm{1.4\,GHz}}$ & $L_{\rm{4.85\,GHz}}$ & $\log R_{\rm X}$  \\
yr & yr &  & erg\,s$^{-1}$ & \\ 
\hline
 $<$18 & $>$8.1 & $>$23 & $7\,10^{40}$ & --1.6 \\ 
\hline                      
\hline
\end{tabular}
\end{table}

\section{Archival optical and IR observations}

\subsection{Long-term optical and IR photometry}

To constrain the onset of the radio outburst that must have happened between 1999.96 (FIRST low-state) and 2017.96 (VLASS high-state), we have inspected archival optical and IR photometric databases in search for high-amplitude variability (from either the accretion disk which could have triggered the change in jet emission, or directly from the increase in non-thermal jet emission itself if extending into the IR-optical regime).  Data prior to 2017.96 are of special interest in constraining the rise phase. Several photometric surveys cover these epochs. 

\paragraph{ASAS-SN.} SDSS\,J1105+1452 has been frequently observed in the course of the All-Sky Automated Survey for Supernovae \citep[ASAS-SN;][]{Shappee2014, Kochanek2017} since 2014\footnote{\url{https://asas-sn.osu.edu/}} at a cadence of several days and with gaps of 3--4 months each year when the object is too close to the sun. 
In Fig.~\ref{fig:MWL-longlight}, we show the Johnson-V and Sloan-g band light curve using co-added data (3 frames per epoch) and image subtraction, including all data points with photometric uncertainties $\leq\pm0.10$ mag. Magnitudes are the integrated magnitudes within an aperture of 16$''$ and include the host galaxy contribution.  

The median g-band magnitude is 17.4,
and the median V-band magnitude 17.09. 
SDSS\,J1105+1452 does not show any systematic changes that could correspond to the factor $\sim$20 increase in the radio regime. 

\paragraph{ZTF.}
SDSSJ1105+1452 has also been observed with the Zwicky Transient Factory (ZTF) since 2018 in the three filter bands r, g, and i 
\citep{Belm2019, Masci2019}. 
We have retrieved the publicly available psf magnitudes (between MJD 55203--57054){\footnote{\url{https://irsa.ipac.caltech.edu/cgi-bin/Gator/nph-scan?mission=irsa&submit=Select&projshort=ZTF} accessed 2026 January 14}}, then filtered out data taken under unfavorable conditions, 
and inspected the remaining ones for any high-amplitude
variability that could represent an outburst in the ZTF bands. None is found, consistent with the conclusions drawn from the ASAS-SN light curve.

\paragraph{Catalina sky survey.} The Catalina survey \citep{Drake2009} has occasionally covered the position of SDSS\,J1105+1452 since the year 2007. The light curve was inspected and no systematic change in the optical magnitude was found. 

\paragraph{SDSS photometry.} SDSS\,J1105+1452 was observed twice with SDSS \citep{York2000}. 
The spectrum was taken on 2005 January 7.
The photometry was carried out earlier on 2003 January 28.  This represents an early optical photometric data set to be inspected for a change in the optical band that could signal the onset of the radio outburst. The integrated SDSS g magnitude 
of 17.7$\pm0.01$ 
\citep{Adelman-McCarthy2009} agrees well with the average ASAS-SN g magnitude of 17.4$\pm{0.2}$. 
The same agreement is found when comparing the psf magnitudes of the core. The SDSS psf g-band magnitude of 18.6 \citep{Anderson2007} is consistent with the average ZTF g-band magnitude of 18.3$\pm{0.1}$. 
This result shows that a major change in optical emission is absent. 

\paragraph{POSS-I and II.} Photometry of SDSS\,J1105+1452 \citep{Monet2003} was also performed in the course of POSS-I (Palomar Observatory Sky Survey; 1949--1965) and POSS-II (1985--2000). While these observations precede the FIRST low-state data point by up to $\sim$50 yr, they still give an impression on past variability of SDSS\,J1105+1452{\footnote{Note that the exact observation dates of any object  
during POSS-I and II are not known \cite{Monet2003}.}} which is very high in the blue band.
During POSS-I, $m_{\rm O,I} = 18.10$  and during POSS-II, $m_{\rm J,II} = 15.78$ with an uncertainty of $\sim$0.25 mag,
or $m_{\rm O,II} = 16.02$ after a correction owing to the different plate emulsions, O and J, used during the two surveys, respectively \citep{Monet2003}. This large magnitude difference corresponds to a factor $\geq 6.8$ flux  variability, and given only two available  epochs, the total amplitude can have well been much higher. 

Given these available constraints, we cannot tell, if this high-amplitude variability was the trigger event for the (delayed) radio turn-on, or was an earlier unrelated event.
In any case it does show that SDSS\,J1105+1452 is capable of undergoing high-amplitude optical continuum variability. 

\paragraph{WISE and NeoWISE.} In the IR, SDSS\,J1105$+$1452 was observed with the Wide-field Infrared Survey Explorer \citep[{WISE,} ][]{wise} in the bands W1--W4 centered on $3.4\mu$m, $4.6\mu$m, $12\mu$m, and $22\mu$m, respectively.
Observations taken during WISE and the more recent NEOWISE survey \citep{neowise} between 2010 and 2024 were retrieved from the NASA/IPAC Infrared Science Archive\footnote{\url{irsa.ipac.caltech.edu} accessed 2025 Aug 2}. 
Measurements with lower photometric quality were excluded (GKK25). 
The only variability seen in the IR bands is a mild decrease in flux density shown in Fig.~\ref{fig:MWL-longlight}. 

\subsection{Optical image} 
\begin{figure}[h!]
   \centering
   \includegraphics[width=8.0cm,
   clip]{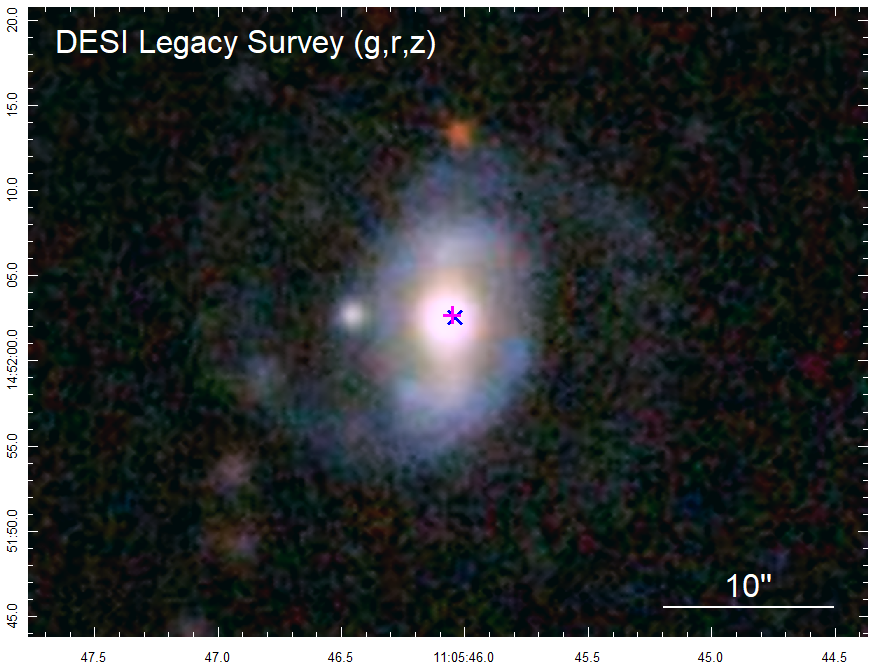}
      \caption{Multi-color DESI image of the barred spiral galaxy SDSS\,J1105$+$1452.
      The radio emission coincides with the center of the galaxy (blue cross: FIRST low-state, magenta plus: VLASS high-state).
      }
         \label{fig:DESIimage}
   \end{figure}

The SDSS image\footnote{\url{skyserver.sdss.org/dr14/en/tools/explore/summary.aspx?} accessed 2025 Oct 1} of SDSS\,J1105+1452 shows a spiral galaxy 
and an additional point source at $\sim$6$''$ from the galaxy center.
The DESI legacy survey image{\footnote{\url{https://www.legacysurvey.org/}}} \citep{Dey2019} goes deeper and reveals the outer spiral arms and additional sources in the field (Fig. \ref{fig:DESIimage}). 
The low-state and high-state radio emission \citep{Helfand_first, Gordon2021} arises from the center of the galaxy and is not associated with the off-nuclear point sources (Fig. \ref{fig:DESIimage}; where the VLASS radio position is the most accurate high-state position). The FIRST low-state and VLASS high-state positional uncertainties are 1$''$ and 0.5$''$, respectively. VLASS agrees with the optical position of SDSS\,J1105+1452 within 0.3$''$ \citep[Gaia DR3;][]{Gaia2023, Gavras2023}.  

\section{SMBH mass estimates, Eddington ratio and NLS1 classification}

\subsection{SMBH mass and Eddington ratio}

In the course of several SDSS large-sample studies, the SMBH mass of SDSS\,J1105+1452 was estimated employing the 5100$\AA$ continuum luminosity, the FWHM of H$\beta$, and broad-line region (BLR)--SMBH single-epoch scaling relations. This approach consistently resulted in $M_{\rm BH} < 10^7$ M$_\odot$ \citep{Sun2015, Rakshit2017, Paliya2024}. Given that the optical continuum emission may have a significant contribution from non-thermal 
jet emission, we alternatively use the H$\beta$ luminosity to 
derive the SMBH mass \citep{Vestergaard2006}. With $L(H\beta)=10^{41.40}$ erg\,s$^{-1}$  \citep{Viswanath2019}\citep[see also][]{Oh2015} 
and 
\begin{equation}
    M_{\rm BH} = 10^{6.67}~~ \left(\frac{L_{\rm H\beta}} {10^{42}\, \rm{erg\,s}^{-1}}\right)^{0.63} ~~\left(\frac{\rm{FWHM(H\beta)}} {1000\, \rm{km\,s}^{-1}}\right)^{2} ~ M_{\odot}
\end{equation}
one obtains $M_{\rm BH} = 4.4 \times 10^{6}$ M$_{\odot}$. 
With $L_{\rm Edd} = 1.3 \times 10^{38}$ $M_{\rm BH}$/M$_{\odot}$ erg\,s$^{-1}$ and $L_{\rm bol} \simeq 10 \times L_{\rm X} = 1.1 \times 10^{44}$ erg\,s$^{-1}$ \citep{Elvis1994, Grupe2010} this then gives an Eddington ratio $\lambda_{\rm Edd}$ = $L_{\rm bol}$/$L_{\rm Edd}$ = 0.2. 

\citet{Vestergaard2006} estimate an uncertainty of 0.5 dex for single-epoch SMBH mass values. Within this uncertainty, the SMBH mass alternatively derived from H$\alpha$ width and luminosity \citep{Rakshit2017} measurements and applying one of the single-epoch relations for H$\alpha$ \citep{DallaBonta2025} is consistent with the H$\beta$-based value. We prefer the H$\beta$ relation because of the extra uncertainty in deblending H$\alpha$ from [N\,{\sc ii}]. 

Alternatively, the SMBH mass can be measured from the host galaxy stellar velocity dispersion $\sigma_*$. 
Using $\sigma_*$ = 82 km\,s$^{-1}$ \citep{Sun2015} and the $M_{\rm BH}-\sigma_*$ relation of non-elliptical galaxies \citep{Gultekin2009},
\begin{equation}
    \log M_{\rm BH} = 8.01 + 4.05\,\log \left(\frac{\sigma_*} {200\, \rm{km\,s}^{-1}}\right)
\end{equation}
then gives $M_{\rm BH} = 2.8 \times 10^{6}$ M$_{\odot}$ and $\lambda_{\rm Edd}$ = 0.3.

\subsection{NLS1 classification}

The optical spectrum of SDSS\,J1105+1452 fulfills all three classification criteria of a NLS1 galaxy (Sect. 1). 
Alternatively, in the main sequence classification scheme of \citet{Marziani2018}, SDSS\,J1105+1452 is in the regime of (extreme) population A sources based on its strong FeII emission and FWHM(H$\beta$). This regime corresponds to NLS1 galaxies. 

Further properties of SDSS\,J1105+1452, the SMBH mass, Eddington ratio, and the soft X-ray spectrum, are consistent with its optical NLS1 classification, demonstrating that its NLS1 nature is very well established. 
This is important because variable radio emission, radio-loudness, and optical changing-look events are all relatively rare in NLS1 galaxies and therefore the NLS1 classification of SDSS\,J1105+1452 and its SMBH mass have to be carefully evaluated. 

\section{Discussion}
SDSS\,J1105+1452 is rare in its high amplitude of radio variability and unique in its outburst duration.
Here, we first discuss constraints from the new MWL observations and then turn to outburst scenarios.

\subsection{X-ray spectrum}

The X-ray spectrum is relatively steep with a photon index of $\Gamma_{\rm X}$ = 2.5, suggesting the presence of a soft X-ray spectral component frequently observed in NLS1 galaxies \citep{Puchnarewicz1992, Grupe2010}. Indeed, the residuals we see in the X-ray spectral fit could indicate the intersection of two components, however SDSS\,J1105+1452 is too faint in X-rays for Swift to allow the fit of two-component models. The fact that the X-ray emission was brighter in the soft (0.1--2.4) keV ROSAT band is consistent with variable accretion-disk related emission as well.
The soft X-ray spectrum is well consistent with the NLS1 nature of SDSS\,J1105+1452. 

\subsection{Extinction and absorption}
X-ray and optical observations can be used to measure the absorption and extinction along our line of sight (l.o.s.).
This information is important for the evaluation of turn-on scenarios, and also for the radio-loudness measurement of SDSS\,J1105+1452. The radio-loudness index \citep{Kellermann1989} is only well defined for unabsorbed AGN. 

The galaxy has a blue continuum and is detected in the UV, implying a lack of heavy extinction.
The Balmer decrement of the BLR, the flux ratio $f(H\alpha)/f(H\beta)=3.3$,  
is close to the case B recombination value \citep{Osterbrock1989} and implies that there is little dust extinction along our l.o.s.
Assuming an intrinsic value of $f(H\alpha)/f(H\beta)=3.1$ (often used to represent the intrinsic emission) and a Galactic reddening law, we obtain a reddening of E(B--V) = 0.05. 
Using the case B value for the Balmer decrement of 2.85 \citep{Osterbrock1989, Gaskell2017} gives E(B$-$V) = 0.12. 

A Galactic gas/dust composition then implies a column density of cold hydrogen of $N_{\rm H} = 1.79 \times 10^{21} \times A_{\rm V}$ \citep{Predehl1995}, where $A_{\rm V} = 3.1 \times E(B-V)$, and in the case of SDSS\,J1105+1452 $N_{\rm H} = 2.8 \times 10^{20}$ cm$^{-2}$.
This value is an upper limit since the Balmer decrement can also be affected by optical depth effects intrinsic to the BLR.

The X-ray spectral fits do not require excess absorption either, and only allow for at most a small amount of intrinsic absorption, $<3.1 \times 10^{20}$ cm$^{-2}$.

\subsection{Constraints on a starburst contribution}

The starburst contribution in the IR--optical regime can be estimated in several different ways. It is important for assessing the broadband properties and evolutionary state of SDSS\,J1105+1452 in general, and its variability characteristics in particular. First, the WISE IR colors locate SDSS\,J1105+1452 in the AGN regime of the IR color-color diagram \citep[GKK25;][their Fig. 26]{jarrett2011}. Second, the faintness of [O\,{\sc ii}]3727 vs [O\,{\sc iii}]5007 in the SDSS spectrum implies that narrow emission-line ratios are not dominated by star formation. Third,  
converting the star formation rate (SFR) measured from the MWL SED, of only 2 M$_\odot$/yr \citep{Barrows2021}, into the associated 1.4 GHz radio emission \citep{Hopkins2003}, we predict a luminosity $L_{\rm 1.4\,GHz,\,SF}  = 3.6 \times 10^{21}$ W/Hz. This value is 15 times {\em lower} than the actually observed FIRST low-state radio emission. 
It implies that the emission arises from AGN activity rather than star formation. 

\subsection{SED}

\begin{figure}
   \centering
   \includegraphics[clip, trim=0.7cm 1.1cm 3.3cm 0.4cm, angle=-90, width=8.4cm]{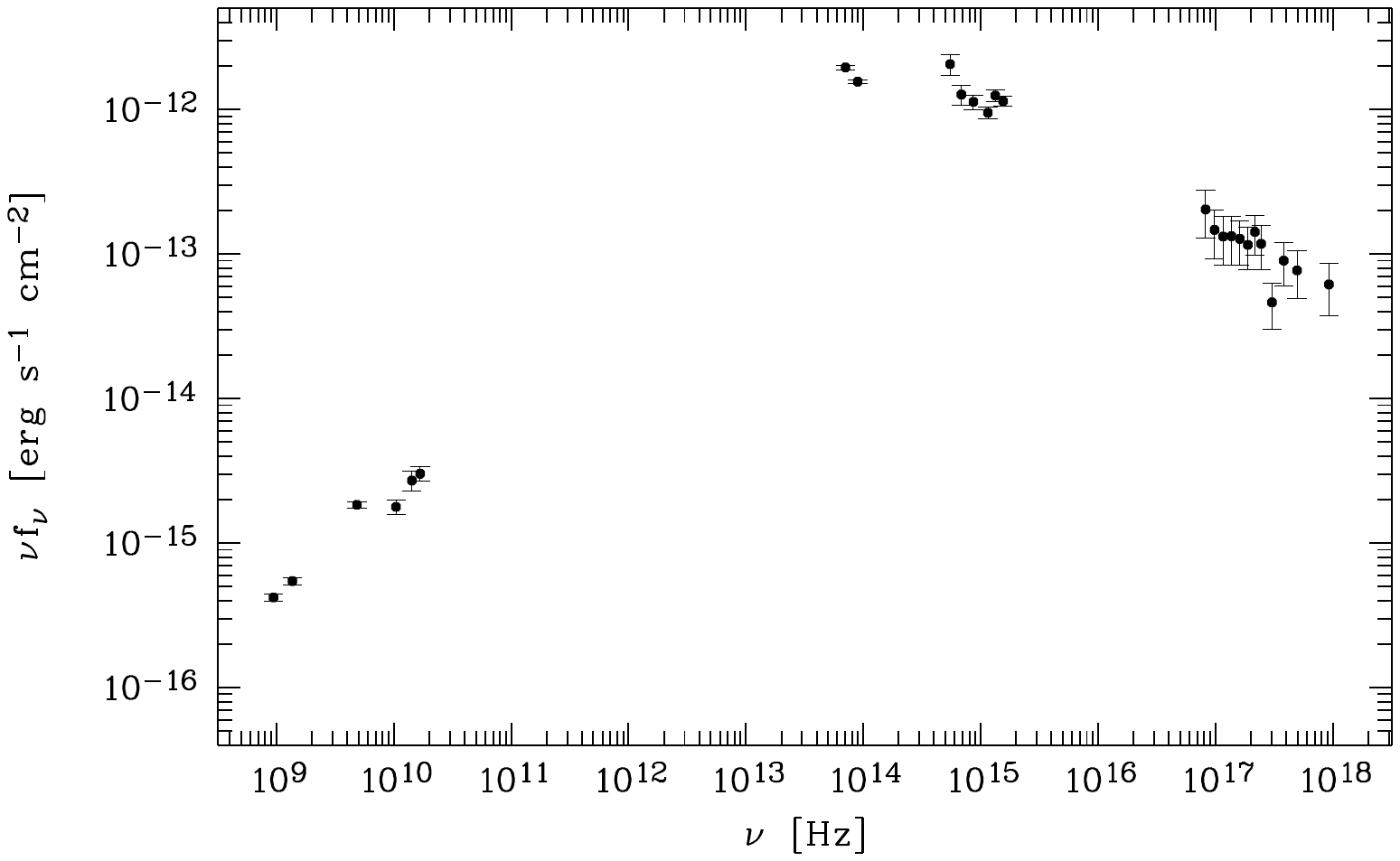}
      \caption{Broadband SED of SDSS\,J1105+1452 from radio to X-rays based on data from RACS, Effelsberg, NeoWISE, Swift UVOT, and Swift XRT.
      }
         \label{fig:broad-SED}
   \end{figure}

\begin{figure}
   \centering
   \includegraphics[clip, trim=0.7cm 1.4cm 3.5cm 0.02cm, angle=-90, width=8.5cm]{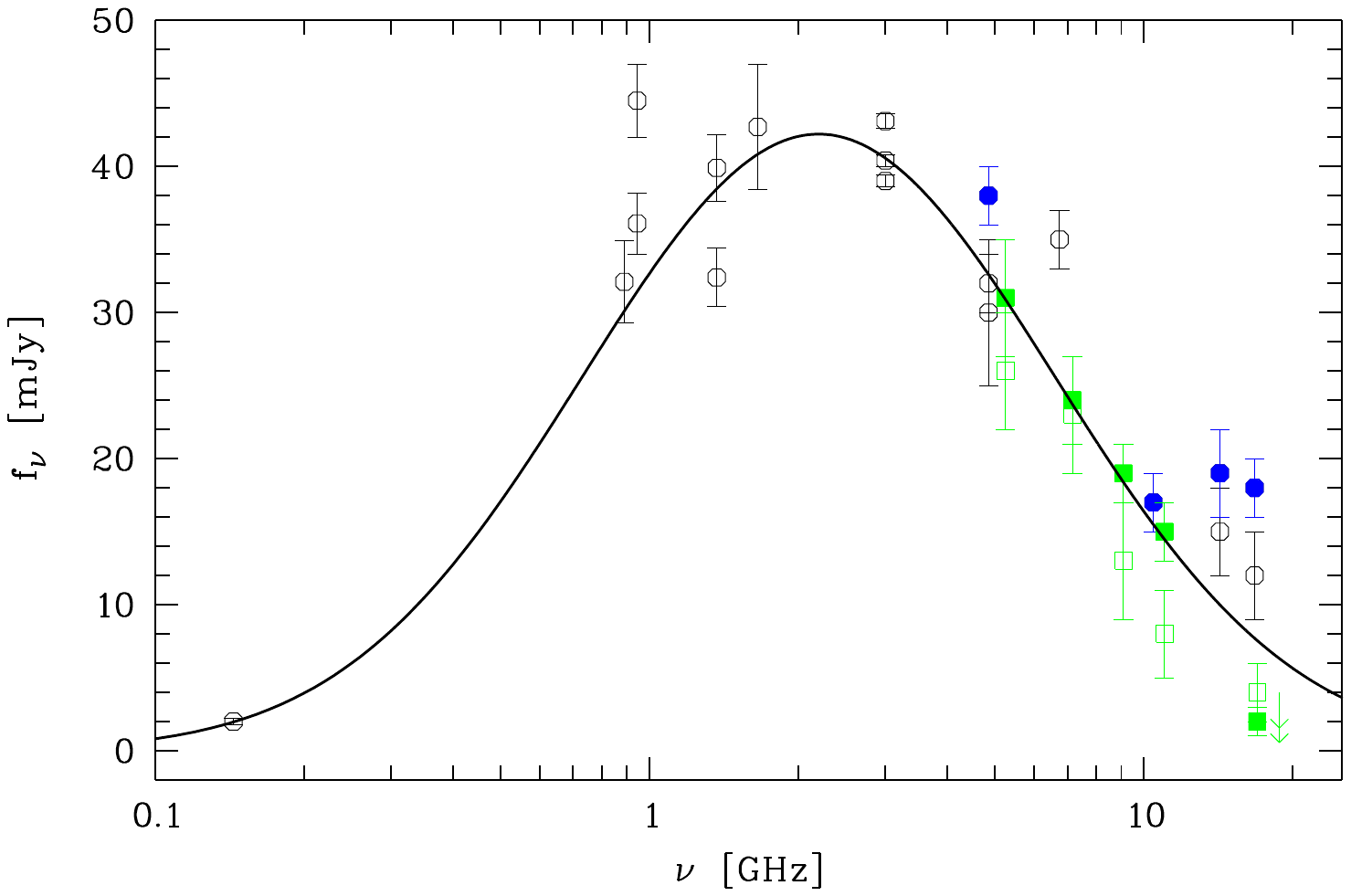}
      \caption{High-state radio SED of SDSS\,J1105+1452. Simultaneous radio observations taken in September 2025 with the Effelsberg 100m telescope are marked with filled blue circles, simultaneous ATCA measurements are shown with open green squares (December 2025) and filled green squares (January 2026). Other flux density measurements are shown with open circles. The best-fit log-space parabola is shown with a black line.
}
         \label{fig:radio-SED}
   \end{figure}

The broadband SED shown in Fig.~\ref{fig:broad-SED} is based on the simultaneous Effelsberg measurements, the RACS observation from 2024, the latest NeoWISE measurements, the Swift UVOT observation of November 2025, and the merged Swift XRT spectrum. 
The 0.8--20 GHz radio spectrum (Fig.~\ref{fig:radio-SED}) is broadly similar to gigahertz-peaked-spectrum (GPS) radio sources \citep{ODea1998}. These are interpreted as young radio jets less than $\sim$1000s years old. 
Two main mechanisms are thought to be responsible for the spectral  turnover in GPS sources \citep[see reviews by][]{ODea1998, ODea2021},
synchrotron self absorption (SSA) and free-free absorption (FFA).
If SSA, the turnover frequency depends on internal properties (e.g., the magnetic field and electron density). An
anticorrelation between source size and turnover frequency is observed
\citep{ODea1997}.
If FFA, the turnover frequency depends on properties of the external medium (e.g., the density of thermal electrons) the jet is interacting with \citep{Bicknell2018} such as a BLR or narrow-line region (NLR) cloud.
The spectra of GPS sources decline steeply toward both lower frequencies (the optically thick part of the  SED) and higher frequencies (the optically thin part).
The majority of GPS sources shows little radio variability \citep{ODea2021} and is hosted by large elliptical galaxies with old stellar populations \citep{deVries2000}, but a fraction of NLS1 galaxies is known to host compact radio sources with peaked spectra as well.

The turnover frequency of SDSS\,J1105+1452 is clearly 
located at low frequencies. 
To determine the turnover value, a parabola  
of the form $\log y = c\,(\log x)^{2} +b\,\log x + a$ was fit to the SED employing \texttt{RadioSED}{\footnote{\url{https://github.com/ekerrison/RadioSED}}} \citep{Kerrison2024}, where $x$ is the frequency in MHz and $y$ is the flux density in Jy. The SED is well fit by this function, with coefficients $a=-11.94{_{-0.25}^{+0.23}}$, 
$b=6.35{_{-0.14}^{+0.15}}$, and $c=-0.95{_{-0.02}^{+0.02}}$ (Fig. \ref{fig:radio-SED}). 
This then implies a turn-over frequency of 2.1 GHz.

At a turnover frequency of 2 GHz, the turnover-size relation \citep{ODea1998} would imply a jet size of $\sim$0.1--0.2 kpc of SDSS\,J1105+1452.  
On the other hand, a very young jet only several years old (as observed), or $\sim$10 pc in size and launched for the first time,
would show a much higher turnover frequency $>$10 GHz within this specific scenario \citep{ODea1998, Nyland2020}. 
However, multiple parameters could affect the SED shape of SDSS\,J1105+1452. It could be peaked at low(er) frequency because an event (like a change in accretion rate) triggered a wider-angle jet or outflow,
or because the magnetic field of SDSS\,J1105+1452 was lower than GPS sources as a class,
thus shifting the Synchrotron peak to lower frequencies.
The spatial resolution of VLASS, 2.5$''$, limits the linear size of the jet of SDSS\,J1105+1452 to $<5.4$ kpc.

We come back to variants of SSA and FFA scenarios in Sect. 7.6. We also note that in the case of exceptional new transients like SDSS\,J1105+1452, the oberved low-frequency decline does not need to be associated with any absorption processes, but could alternatively represent the intrinsic energy distribution of the ejected electron population, with a lack of low-energy electrons.

Significant radio variability is absent. Small deviations at a roughly $\lesssim10-20\%$ level 
could be due to interstellar scintillation \citep{Hancock2019}, a small extent of the radio emission (the bulk is pointlike) in combination with the different spatial resolutions of the surveys and instruments, and/or a small underestimate of the measurement uncertainties.

\subsection{Timing of the turn-on}

Radio observations place the epoch of the rise phase between 1999.96 (the FIRST low-state) and 2017.96 (the VLASS high-state). 
It is plausible to assume that the turn-on also affected the IR/optical band either by accretion disk variability that triggered the jet change, and/or by the synchrotron emission itself extending into the IR/optical band.
Therefore, we have inspected available MWL surveys in search for a factor $\sim$20 (or any sytematic{\footnote{Note that in the absence of detailed disk-jet modelling it is a priori unknown whether an {\em increase} or {\em decrease} in accretion rate would trigger an increase in radio jet emission, or whether any other re-configuration in the disk structure and/or emission not directly linked to the accretion rate affected magnetic flux accumulation. Further, it is possible that there is a time delay between the accretion-disk change observed in optical--X-ray emission, and the detection of radio-jet emission.}}) change in the flux level. We did not detect any. In the IR, WISE dates back to 2010. In the optical, ASAS-SN dates back to 2014, the Catalina survey back to 2007, and SDSS photometry back to 2003. 
Therefore, if the radio turn-on was accompanied by a significant optical/IR change, the trigger likely occurred before 2003.  

Is there any possibility that the turn-on could have happened later (i.e., between 2003 and 2017.96), but we did not detect it in the IR and optical bands? 
First, one might speculate that absorption and extinction affected the 
X-ray, IR and optical bands, making any turn-on undetectable in those bands. This scenario can be tested by measuring the line-of-sight extinction from the BLR Balmer decrement, and the X-ray absorption from the spectral fit. No significant absorption is detected either way (Sect. 7.2), and we therefore reject this scenario. 
Next, we raise the question of whether a change in emission could have been hidden in the IR-optical by a strong star formation component, such that its change was present but remained undetected. However, as discussed above (Sect. 7.3), the IR emission is dominated by the AGN.  

Finally, we note that owing to the lack of dense X-ray coverage, one cannot yet
exclude the possibility that an earlier, pre-1999.96, change of the inner accretion disk, e.g. as traced by the higher X-ray state observed during the
RASS in 1990 and/or by the higher blue magnitude during the POSS, already triggered the jet launch but a significant rise in radio
emission only occurred years later (between 1999.96 and 2017.96) after the jet encountered and shocked high-density
material.

\subsection{Outburst scenarios}

The coincidence of the radio source with the center of the galaxy strongly suggests a link of the radio outburst to nuclear activity. However, among NLS1s and other types of AGN the radio properties of SDSS\,J1105+1452 are unique, and none has shown such a high-amplitude and long-lasting radio outburst at low frequencies before. 
The different scenarios can be sorted into different categories: (1) Outbursting compact sources unrelated to the SMBH environment, like radio supernovae, are not discussed here further since the radio-brightest ones known are still factors $\sim$100 less luminous than SDSS\,J1105+1452 \citep[e.g.,][]{Zhang2022}.  (2) Extrinsic variability caused by gravitational lensing or by absorption, or magnification due to beaming.
(3) Intrinsic variability triggered in the SMBH environment. There are three main processes thought to affect jet launching: First, accretion-disk variability including catastrophic events like TDEs.
Second, (a change in) black hole spin. A spin change is difficult to achieve on short timescales, except for a merger with a second SMBH. Third, changes in magnetic flux accumulation near the horizon.
Below, we discuss these different scenarios using constraints from the many MWL observations of SDSS\,J1105+1452.

\subsubsection{Gravitational lensing} 
Lensing of a compact radio knot can enhance the radio emission if a lens at a suitable location along the l.o.s.\ exists. We have carefully inspected the host galaxy image and optical spectrum of SDSS\,J1105$+$1452. Neither do we recognize evidence for any foreground galaxy in the image, nor is there a second emission-line system with a different redshift in the optical spectrum that could indicate the presence of a lense. 
 
\subsubsection{Absorption variability}
H\,{\sc i} absorption variability. This scenario considers absorption variability, and assumes that the high-state rather represents the `normal' state, and the low-state was due to an extreme absorption event. 
In this scenario, the early FIRST and NVSS  observations at 1.4 GHz would have been severely affected by H\,{\sc i} 21 cm absorption. For this mechanism to work, one has to assume a compact jet shielded by a BLR cloud or torus clump along our l.o.s.
The radio flux is diminished as $S$(1.4, low) = $S$(1.4, high) $e^{-\tau}$, where $\tau$ is the optical depth.
$\tau$ depends on the gas column density and spin temperature $T_{\rm s}$. 
At typical BLR column densities $>$10$^{23-24}$ cm$^{-2}$ \citep{Netzer2013, Panda2021}, the optical depth can be significant. 
However, at the redshift of SDSS\,J1105+1452, 21\,cm absorption would be centered at 1.25 GHz and could extend into the FIRST and NVSS band only at high BLR velocity dispersion of a few 10000 km/s, not expected for the wings of the BLR profile of a NLS1 galaxy. 
Further, the 4.85 GHz band should not be affected, but the GB6 data still show a significantly lower radio flux than the recent Effelsberg and ATCA observations at the same frequency. 
Further, there is no other evidence for strong absorption along our l.o.s. For instance, the RASS detection in the soft X-ray band (several years before the NVSS non-detection) implies the absence of such high gas column densities along our l.o.s. at that epoch. 

FFA by an extended or compact ionized medium. In this scenario, a steady jet that has formed in the past centuries or millenia traversing ionized material (absent along our l.o.s. to the accretion-disk region; see MWL absorption constraints in Sect. 7.2) is subject to FFA. The free–free optical depth is given by 
$\tau_{\rm ff} \simeq 0.082\,(T/{\rm K})^{-1.35} \, (\nu/{\rm GHz})^{-2.1} \,EM/{\rm (pc\,cm^{-6})}$ \citep{Mezger1967}. 
A decrease in emission measure $EM$ could then explain the rise at low frequencies. However, this scenario would require a relatively sharp edge of the absorber, and across a significant part of the width of a jet or lobe, to explain the rapid rise within only decades, and it predicts strong changes in spectra and fluxes as the absorbing medium clears, in addition to light travel-time delays across a jet with an opening angle of few ly to a few dozen ly, that are not observed. If, instead, only a small fraction of the whole jet broke free, the high amplitude of variability and high radio luminosity in the radio-loud regime would be surprising, and it would imply an even higher total amplitude of variability.
Sharp gas-cloud boundaries are more naturally encountered in the dense core environment, when a newly launched, more confined jet passes through an ionized high-density cloud. For instance, at $v \simeq c$, a new jet component could reach a coronal-line region (CLR) cloud after 1 yr. At a typical CLR cloud column density of $N_{\rm H} = 10^{22}$ cm$^{-2}$ and a density of 10$^6$ cm$^{-3}$ \citep{Ferguson1997}, the whole cloud of dimension 10$^{16}$ cm can be crossed in days, well consistent with the rise time of $<$18 yr. The high-state duration of $>$8 yr then implies that no other dense material was encountered; possible for a low covering fraction of the CLR.
However, at a typical jet opening angle of $\sim$1 ly and a cloud dimension of $\sim$10$^{-2}$ ly, only a small fraction of the jet would be subject to FFA, implying a giant total amplitude of variability to raise the integrated luminosity by a factor of 23. 
Further, the question regarding the initial trigger mechanism of such a newly launched jet component remains unanswered in this scenario.      

SSA in the core region.  One can speculate that jet activity is intermittent \citep{Reynolds1997, Kunert2011} and a new jet component of SDSS\,J1105+1452 was recently launched (due to an unspecified trigger mechanism). If initially dense and compact, the blob would experience SSA such that the few GHz region was highly optically thick in the 1980s and 1990s
but became optically thin afterwards leading to a shift of the Synchrotron peak to lower frequencies and therefore to a rapid rise in the observed low-frequency radio emission.
The question is then raised, why the radio spectrum would evolve very rapidly, from $\sim$10 GHz-peaked to 2 GHz-peaked within years, whereas this evolution usually takes centuries to millennia in known GPS sources \citep{ODea1998}. Rapid expansion in the absence of a dense confining medium could be a possibility (see also Sect. 7.4). 
However, first, if the overall high-frequency $\sim$10 GHz radio emission only increased mildly due to the new blob, then we would still expect the emission from older optically thin components to extend into the 1--2 GHz band, effectively leading to only mild variability at low frequency (which is not the case).  If instead, the 10 GHz flux increased by a high factor
 we would still require high-amplitude variability in radio flux in the first place, and the question about the trigger mechanism remains. 

 \subsubsection{Binary SMBH jet swing, or spin flip following coalescence} 
A jet crossing the observer's l.o.s. due to precession caused by the presence of a binary SMBH (or by other jet processes) could lead to a radio outburst due to temporary Doppler boosting. 
Alternatively, binary SMBH coalescence could lead to a newly formed inner accretion disk around the single SMBH different from the pre-coalescence circumbinary disk \citep{DOrazio2013} on a timescale of 
$\sim7\,(1+z)\,(M/10^6\rm{M}_{\odot})$ yr \citep{Milosavljevic2005}
with different jet properties, or it could lead to a spin flip of the newly formed single SMBH \citep{Campanelli2007} and therefore launch a new jet in a new direction. 
This latter scenario is attractive because high-amplitude variability, a drop and re-rise, is only expected for the inner (X-ray emitting) disk, the evolution is rapid for low-mass SMBHs as in NLS1 galaxies, and SMBH coalescences are the most efficient known mechanism to change SMBH spin on human timescales. 
However, there is no evidence based on the DESI host galaxy image that SDSS\,J1105$+$1452 has undergone a recent merger. The extended structures visible in the image are just the outer extension of the inner spiral arms rather than tidal tails. Without any further positive evidence of an ongoing or past merger, we do not discuss this rare scenario any further. 

\subsubsection{TDE} 

With the X-ray discovery of the first Tidal Disruption Events (TDEs), a dedicated search for associated radio (jet) emission  \citep{KomossaDahlem2001, Komossa2002} was carried out. More sensitive recent radio observations have provided a high detection fraction of TDEs.
The radio emission of the known TDEs, interpreted as jets and/or outflows, typically peaked $\sim1-3$\,yr after the optical or X-ray event, and then declined \citep{Cendes2024, Anumarlapudi2024}. This is markedly different from SDSS\,J1105$+$1452 which remains radio bright and at roughly constant emission level for at least $8$\,yr implying a long-lasting constant power input. 
Slow circularization of the stellar debris \citep{Piran2015, Lin2022} and/or rare disruption events of giant stars \citep{Guillochon2015, Readhead2024} might provide the required long-lasting near-constant power input 
but they have not yet been detected and further explored in the radio regime and SDSS\,J1105+1452 would represent the first such case. 
The longer the radio outburst lasts at constant emission levels, the more challenging it will be to understand the outburst within this framework. 
 Given the rarity of (giant) star disruptions and given the fact that SDSS\,J1105+1452 was already a (radio-)AGN before the outburst,
we do not favor a TDE interpretation. We cannot entirely rule it out either, and will re-evaluate this scenario as our radio monitoring continues.

 \subsubsection{Activation of AGN activity for the first time} 
 NLS1 galaxies are thought to be young AGN in an early stage of evolution \citep{Mathur2000, Grupe2004}. Therefore, is it conceivable that we may have witnessed the very first turn on of accretion and AGN (jet) activity? This scenario can be rejected because the SDSS spectrum of 2005 already shows a classical NLR (traced by transitions like [Ne\,{\sc iii}]3798, [O\,{\sc iii}]5007, [O\,{\sc i}]6300, and [S\,{\sc ii}]6716,31) implying AGN activity during at least the last 1000s of years given light-travel time constraints ($t_{\rm{light}} \sim$ 100s--1000s of years) and recombination timescale arguments ($t_{\rm{rec}} \sim$ 1000--10000s of years).  
 Further, the FIRST low-state radio emission implies a pre-existing jet (Sect. 7.3). 
 
\subsubsection{Blazar variability}
Variability like that observed in blazars, due to Doppler boosting in a jet closely aligned with our l.o.s., was disfavored by GKK25. 
The variability pattern of SDSS\,J1105+1452 is very different from that of classical blazars that flare on timescales of days to months, but do not remain in high-state and at near-constant emission levels for years \citep{Aller1985, Urry1996, Marscher2016}. The steep optically thin part of the radio spectrum beyond a few GHz (Sect. 4, Fig. \ref{fig:radio-SED}) is different from blazars as well.

\subsubsection{Change in accretion rate}

In the last decade, a significant number of optical changing-look 
AGN have been identified. These are characterized by high-amplitude continuum and broad emission line variability such that they change their optical spectroscopic Seyfert classification between type 1 and type 2 (or type 1.9, 1.8) and vice versa \citep[review by][]{Komossa2026}. 
These changes occur on timescales of months to decades. Some systems turn on or off over years 
\citep[][]{Denney2014}, others show fast changes within months and/or repeatedly transition between type 1 and type 2 \citep{Alloin1986, Popovic2023, Ochmann2024}. While the first changing-look events have been recognized in the 1960--80s \citep{Andrillat1968, Alloin1985, Kollatschny1985}, larger numbers have been identified in recent years thanks to large-area optical sky surveys \citep{Green2022, Guo2024, Shen2025}, and they have been detected not only in Seyfert galaxies but also in quasars \citep{LaMassa2015, Wang2018, Guo2025}.

The high-amplitude continuum variability of changing-look AGN implies a change in accretion-disk emission, but the physical drivers behind rapid changes in the accretion rate much faster than the viscous timescale have remained largely unknown, and a variety of different models have been explored in recent years \citep[e.g., ][]{Grupe2015, Stern2018, Sniegowska2020, Wang2024x, LiCao2025, Kaaz2026}, including variants where magnetic fields play an important role \citep{DexterBegelman2019, Scepi2021, Laha2022, Wu2023, Cao2023}.   

A few changing-look AGN with dedicated or archival radio follow-ups have shown clear changes in their radio emission \citep{Koay2016, Meyer2025, Jana2025, Birmingham2025}. 
The majority of optical changing-look events has so far been identified in BLQ1s and BLS1s. In recent years, an intense search for similar optical events in NLS1 galaxies has been ongoing, and only a few cases have been identified so far \citep{Frederick2019, Hon2022, Xu2024, Komossa2024, Wang2026}. 
SDSS\,J1105+1452 is the first case that shows a long-lived radio changing-look. 

We suggest that the radio turn-on event of SDSS\,J1105+1452 was due to a change in accretion rate that triggered a dramatic increase in jet power and thus radio emission. The majority of recent accretion-disk variability models \citep[e.g., ][]{Stern2018, DexterBegelman2019, Sniegowska2020, Wang2024x, LiCao2025, Kaaz2026} do not yet provide immediate predictions for the changes in radio jet emission and evolution.
In particular, the amplitude of optical or X-ray variability (any increase or decrease) does not need to match the amplitude of radio variability (an observed factor 23),
and there can be a time delay between the optical--X-ray change and the increase in radio emission (see also footnote 12).

As an example, we consider the following scenario. A way to reconcile a modest change in the accretion luminosity (e.g., a factor of at least 2 variability suggested by the low-cadence X-ray observations) with an extreme radio brightening (factor $\sim$20) is to allow the jet efficiency to change nonlinearly, controlled by the net poloidal magnetic flux threading the SMBH, $\Phi_{\rm BH}$. 
In Blandford–Znajek–type jets the jet power scales approximately as $P_{\rm jet} \propto \Phi_{\rm BH}^2 \Omega_{\rm H}^2$  \citep{Blandford1977}, 
and GRMHD simulations show that once the inner flow approaches the magnetically arrested disc (MAD) limit, the jet efficiency can increase dramatically at nearly fixed radiative output \citep{Narayan2003, Tchekhovskoy2011, McKinney2012}. Importantly, the mapping from jet power to cm-band radio luminosity is itself nonlinear; empirical X-ray-cavity--jet-power scalings imply $P_{\rm jet} \propto L_{\rm R}^{0.7-0.8}$ \citep{Cavagnolo2010}, so a 20 fold increase in radio luminosity corresponds to only a 6--9 fold increase in kinetic power. 
We emphasize that the cavity-power scaling is used here only to illustrate the nonlinear mapping between radio luminosity and jet power, not as a precise estimator of $P_{\rm jet}$ for this compact NLS1 galaxy.
If $P_{\rm jet} \propto \Phi_{\rm BH}^2$ (at fixed BH spin), the required change in $\Phi_{\rm BH}$  is then only a factor 2.4--3, i.e., a comparatively modest increase in net magnetic field flux can plausibly yield an order-of-magnitude radio transition without invoking an order-of-magnitude change in $\dot{M}$. Such flux growth can occur if the inner disc crosses from a diffusion-dominated to an advection-dominated flux-transport regime, because the inward advection of large-scale field competes with outward turbulent diffusion \citep{Lubow1994}. 
The rarity of such events in NLS1/BLS1 populations can be qualitatively understood if the supply of coherent net vertical flux (i.e., a sustained polarity imbalance) is itself intermittent: many accretion episodes may deliver mixed-polarity fields that cancel before reaching the black hole, whereas only a small subset of episodes advect sufficient net $\Phi_{\rm BH}$  to cross a MAD-like threshold and trigger a high-efficiency 
jet \citep[see also][]{An2026}.
Recent VLBI observations of relativistic jet activity of the radio-quiet AGN Mrk\,110 provide observational support for such episodic jet activation in a radio-quiet/NLS1-like system \citep{WangAn2025}.

In summary, there is ample evidence for a coupling between disk and jet changes in classical AGN \citep{Marscher2002} and Galactic binaries \citep{Fender2004}, 
and the violent and rapid changes in the structure and dynamics of magnetized disks in changing-look events are expected to directly fuel the launch of jets and outflows. 

\section{Summary and conclusions}

We have reported new and first observations across the electromagnetic spectrum of the recently discovered unique long-duration radio outburst of a NLS1 galaxy. Owing to its low redshift, a large number of archival and new MWL observations could be obtained that would have been missing at higher redshift.
The salient results on MWL properties and interpretation of the unprecedented  
radio changing-look event of SDSS\,J1105+1452 can be summarized as follows: 

$\bullet$ Different methods of SMBH mass determination provide a mass in the range (2.8--4.4) $\times$ $10^6$ M$_\odot$ and a high Eddington ratio $\lambda_{\rm Edd}=0.2-0.3$. These values corroborate the optical NLS1 classification. 

$\bullet$ The photon index of the first (0.3--10) keV X-ray spectrum of SDSS\,J1105+1452, $\Gamma_{\rm{X}}=2.5$, shows that the spectrum is steep and consistent with a contribution from accretion-disk emission. 

$\bullet$ The absorption and extinction along our l.o.s., measured from X-ray and optical observations, is low; $E(B-V) \leq 0.05 - 0.1$ (from the optical) and $N_{\rm H} \leq 3.1 \times 10^{20}$ cm$^{-2}$ (from X-ray spectral fits). Therefore, along our l.o.s. significant amounts of dusty or dust-free absorbers are absent. 

$\bullet$ Several different arguments demonstrate that the star formation activity is moderate and the low-state radio emission is still dominated by the AGN. 

$\bullet$ The radio SED, measured for the first time beyond several GHz, declines steeply at high frequencies. The SED is low-frequency peaked.
It is well fit by a phenomenological parabola in log-log space
and turns over at 2.1 GHz. 
When described by a powerlaw, the SED declines with a spectral index 
$\alpha_{\rm{thick}} = 1.6\pm{0.1}$ toward low frequencies.
At 4.85 GHz, $L_{\rm 4.85\,GHz}=7\times10^{40}$ erg s$^{-1}$. 

$\bullet$ Radio observations between December 2017 and January 2026 show that the radio outburst is remarkably long-lived. It has lasted at least 8 years at roughly constant flux density. 

$\bullet$ Inspection of archival optical/IR data 
reveals no systematic high-amplitude change since 2003. 
Therefore, if the radio turn-on was accompanied by a significant optical/IR change, the trigger likely occurred before 2003.
This includes the possibility that an earlier disk change as traced by, e.g., the higher X-ray state observed during the RASS, or the high blue magnitude during the POSS, triggered the jet launch but a significant rise in radio emission only occurred several years later.

$\bullet$ A variety of different outburst scenarios were considered, 
including lensing, variants of absorption variability, the presence of a binary SMBH or a coalescence-driven spin flip, a TDE,  AGN ignition for the first time, 
and accretion-disk variability as in optical changing-look events. 
The majority of them can either be rejected or they are deemed too rare (an SMBH spin flip following binary coalescence, or a giant-star TDE) to warrant further exploration at this stage in the absence of supporting evidence. However, in the era of SKA with ongoing and up-coming deep large-area sky surveys, we expect that all of these transient phenomena and physics will be discovered and explored at some stage. 
We suggest that a change in the accretion rate as observed for instance in optical changing-look AGN powered a long-lived change, a strong increase, in the radio jet emission of SDSS\,J1105+1452. 

Multiple follow-up observations are ongoing and/or are of interest.
Ongoing radio monitoring will follow the evolution of the radio jet and enable a search for any turn-off. This then offers an excellent chance to trace radio (and MWL) turn-off physics. SDSS\,J1105+1452 may even undergo recurrent radio changing-look transitions, as observed in the optical regime in a fraction of the optical changing-look events in broad-line type 1 AGN \citep[e.g.,][]{Wang2024x, Panda2024, Dong2025}. 
High-resolution VLBI imaging will determine the location and compactness of the jet down to parsec scales. 
HI 21cm absorption spectroscopy at the redshift of SDSS\,J1105+1452 will provide the amount of cold hydrogen absorption directly along our l.o.s. toward the radio source. 
New optical spectroscopy will allow us to search for evolution in the emission lines.  
Deep XMM-Newton X-ray spectroscopy will disentangle the soft excess and powerlaw (jet and corona) emission. 
SDSS\,J1105+1452, despite its excellent coverage, still suffered from gaps in MWL data. A dedicated search for new, similar transient events in their rise phase in currently ongoing radio surveys 
will have the benefit of available 
simultaneous dense coverage in optical surveys including LSST \citep[Large Synoptic Survey Telescope;][]{Ivezic2019}.  

While the majority of NLS1 galaxies are constant radio emitters, 
the case of SDSS\,J1105+1452
highlights that some can undergo decade-long dramatic radio changing-look transitions between radio-quiet and radio-loud, and SDSS\,J1105+1452 
may be the prototype of a new time-domain radio AGN class. 
Establishing the physics behind new types of long-lived radio transients provides us with important new insights into the
mechanisms of radio jet formation and evolution under extreme
conditions and their impact on the ISM, and sets the stage
for the future search of such transients in the era of SKA. 


\section*{Acknowledgments}
We would like to thank the Swift team for carrying out the observations we proposed. 
It is our pleasure to thank Krzysztof Stanek for many very helpful discussions about the ASAS-SN data and Emil Lenc for very useful discussions. 
We would like to thank our referee for many useful suggestions. 
SK would like to thank the CAS President's International Fellowship Initiative for partial support. 
SK and RS would like to thank NAOC Beijing for their hospitality. SK would also like to thank Shanghai Observatory for their hospitality. MJH thanks the UK STFC for support [ST/Y001249/1].  
KÉG was supported by HUN-REN, and the NKFIH excellence grant TKP2021-NKTA-64. 
SP is supported by the international Gemini Observatory, a program of NSF NOIRLab, which is managed by the Association of Universities for Research in Astronomy (AURA) under a cooperative agreement with the U.S. National Science Foundation, on behalf of the Gemini partnership of Argentina, Brazil, Canada, Chile, the Republic of Korea, and the United States of America.
DWX is supported by the National Natural Science Foundation
of China under grant number 12273054.
This work is partly based on
data obtained with the 100\,m telescope of the Max-Planck-Institut für Radioastronomie at Effelsberg. 
The Australia Telescope Compact Array is part of the Australia Telescope National Facility (https://ror.org/05qajvd42) which is funded by the Australian Government for operation as a National Facility managed by CSIRO.
We acknowledge the Gomeroi people as the Traditional Owners of the Observatory site.
This work uses data obtained from Inyarrimanha Ilgari Bundara / the CSIRO's Murchison Radio-astronomy Observatory. We acknowledge the Wajarri Yamaji People as the Traditional Owners and native title holders of the Observatory site. CSIRO's ASKAP radio telescope is part of the Australia Telescope National Facility (https://ror.org/05qajvd42). Operation of ASKAP is funded by the Australian Government with support from the National Collaborative Research Infrastructure Strategy. ASKAP uses the resources of the Pawsey Supercomputing Research Centre. Establishment of ASKAP, Inyarrimanha Ilgari Bundara, the CSIRO Murchison Radio-astronomy Observatory and the Pawsey Supercomputing Research Centre are initiatives of the Australian Government, with support from the Government of Western Australia and the Science and Industry Endowment Fund. 
LOFAR is the Low Frequency Array designed and constructed by ASTRON. It has observing, data processing, and data storage facilities in several countries, which are owned by various parties (each with their own funding sources), and which are collectively operated by the LOFAR ERIC under a joint scientific policy. The LOFAR resources have benefited from the following recent major funding sources: CNRS-INSU, Observatoire de Paris and Université d'Orléans, France; BMBF, MIWF-NRW, MPG, Germany; Science Foundation Ireland (SFI), Department of Business, Enterprise and Innovation (DBEI), Ireland; NWO, The Netherlands; The Science and Technology Facilities Council, UK; Ministry of Science and Higher Education, Poland; The Istituto Nazionale di Astrofisica (INAF), Italy.
This research made use of the Dutch national e-infrastructure with support of the SURF Cooperative (e-infra 180169) and the LOFAR e-infra group. The Jülich LOFAR Long Term Archive and the German LOFAR network are both coordinated and operated by the Jülich Supercomputing Centre (JSC), and computing resources on the supercomputer JUWELS at JSC were provided by the Gauss Centre for Supercomputing e.V. (grant CHTB00) through the John von Neumann Institute for Computing (NIC).
This research made use of the University of Hertfordshire high-performance computing facility and the LOFAR-UK computing facility located at the University of Hertfordshire and supported by STFC [ST/P000096/1], and of the Italian LOFAR IT computing infrastructure supported and operated by INAF, and by the Physics Department of Turin university (under an agreement with Consorzio Interuniversitario per la Fisica Spaziale) at the C3S Supercomputing Centre, Italy.
This research is part of the project LOFAR Data Valorization (LDV) [project numbers 2020.031, 2022.033, and 2024.047] of the research programme Computing Time on National Computer Facilities using SPIDER that is (co-)funded by the Dutch Research Council (NWO), hosted by SURF through the call for proposals of Computing Time on National Computer Facilities. 
This research has made use of the NASA/IPAC Extragalactic Database (NED), which is operated by the Jet Propulsion Laboratory, California Institute of Technology, under contract with the National Aeronautics and Space Administration. 
This research has made use of NASA's Astrophysics Data System (ADS) Bibliographic Services.  

\vspace{4mm}
\facilities{Neil Gehrels Swift observatory (XRT and UVOT), Effelsberg 100m telescope, ATCA, ASKAP, and LOFAR.}

\software{HEASoft (\url{https://heasarc.gsfc.nasa.gov/docs/software/heasoft/}) with XSPEC \citep{Arnaud1996}, ESO-MIDAS (\url{https://www.eso.org/sci/software/esomidas/}), XRT Data Analysis Software (XRTDAS) developed under the responsibility
of the ASI Science Data Center (ASDC), Italy (\url{https://www.ssdc.asi.it/}), RadioSED \citep[][\url{https://github.com/ekerrison/RadioSED}]{Kerrison2024}, CASA \citep{Bean2022}, and Miriad \citep{Sault1995}.  
}

\section{Data Availability}
The Swift data of our project are available in the Swift archive at \url{https://swift.gsfc.nasa.gov/archive/}. Effelsberg and ATCA data are available upon reasonable request. Archived ASKAP data can be obtained through the CSIRO ASKAP Science Data Archive, CASDA, at \url{http://data.csiro.au/}. 



\end{document}